\documentclass{aa}

\usepackage{url}
\usepackage[breaklinks=true]{hyperref}
\usepackage{twoopt}
\usepackage[english]{babel}          

\usepackage{natbib}
\bibpunct{(}{)}{;}{a}{}{,} 

\usepackage[utf8]{inputenc}
\usepackage[normalem]{ulem}
\usepackage{siunitx}
\usepackage{amsmath, amssymb}
\usepackage[compatibility=false]{caption}
\usepackage[farskip=0pt]{subfig}

\usepackage{multirow,bigstrut,ctable}
\usepackage[normalem]{ulem}
\usepackage{rotating}
\usepackage[varg]{txfonts} 
\usepackage{threeparttable}

\def \deg         {\text{$^{\circ}$}}
\def \arcmin      {\text{$^\prime$}}
\def \arcsec      {\text{$^{\prime\prime}$}}

\def \mjybeam     {mJy\,beam$^{-1}$}

\newcommand{\Msun}{\text{$\rm M_\odot$}}

\begin{document}

\title{Single- and double-headed odd radio circles in the LOFAR surveys}

\author{F.~De~Gasperin\inst{\ref{ira}} \and M.~Br\"uggen\inst{\ref{ham}} \and 
T.~Pasini\inst{\ref{ira}} \and H.~Andernach\inst{\ref{tur},\ref{mex}} \and A.~Ivleva\inst{\ref{inst:usm}} \and C.~Spingola\inst{\ref{ira}} \and L.~M.~B\"oss\inst{\ref{inst:chicago}} \and 
J.~Callingham\inst{\ref{inst:astron},\ref{inst:amst}} \and
M.~Hardcastle\inst{\ref{inst:hert}} \and
G.~Di~Gennaro\inst{\ref{ira}}}
\authorrunning{F.~de~Gasperin et al.}

\institute{
INAF - Istituto di Radioastronomia, via P. Gobetti 101, 40129, Bologna, Italy \email{fdg@ira.inaf.it} \label{ira}
\and
Hamburger Sternwarte, Universit\"at Hamburg, Gojenbergsweg 112, 21029, Hamburg, Germany \label{ham}
\and
Th\"uringer Landessternwarte, Sternwarte 5, D-07778 Tautenburg, Germany \label{tur}
\and
Permanent address: Depto.\ de Astronom\'{i}a, DCNE, Universidad de Guanajuato,
Callej\'on de Jalisco s/n, C.P. 36023 Guanajuato, GTO, Mexico \label{mex}
\and
Universitäts-Sternwarte, Fakultät für Physik, Ludwig-Maximilians-Universität München, Scheinerstr. 1, 81679 München, Germany\label{inst:usm}
\and
Department of Astronomy and Astrophysics, The University of Chicago, William Eckhart Research Center, 5640 S. Ellis Ave. Chicago, IL 60637\label{inst:chicago}
\and
ASTRON, Netherlands Institute for Radio Astronomy, Oude Hoogeveensedijk 4, Dwingeloo, 7991PD, The Netherlands\label{inst:astron}
\and
Anton Pannekoek Institute for Astronomy, University of Amsterdam, Science Park 904, 1098XH, Amsterdam, The Netherlands\label{inst:amst}
\and
Department of Physics, Astronomy and Mathematics, University of Hertfordshire, College Lane, Hatfield AL10 9AB, UK\label{inst:hert}
}

\date{Received ... / Accepted ...}

\abstract
{Deep radio surveys are now producing catalogs with millions of radio sources. Radio sources can have complex morphologies that depend on both the production mechanisms and the local environment. Recently, an unusual class of circular radio sources named "odd radio circles" (ORCs) were identified. They have sizes of about 1\arcmin, a circular/elliptical shape and appear edge-brightened. Subsequent observations have suggested that this class may comprise a variety of sources. Despite various attempts to explain them, their origin remains unclear.}
{The main goal of this work is to increase the number of known ORCs and derive common characteristics that can help identify the origin of these sources.}
{We searched the LOFAR Two Metre Sky Survey (LoTSS) Data Release 3 (DR3) at 144 MHz for ORCs using a combination of parameter filtering on catalog entries and visual inspection. We then identified possible optical counterparts and derived information such as redshift, physical size, and spectral index using further radio data at 54 and 1400 MHz.}
{We isolated 18 sources with ORC structures. Four of these are double-headed ORCs (ORCs with two rings), and two are new discoveries. We also found five new single-headed ORCs and nine candidate ORCs.}
{With this work we significantly expand the population of known ORCs. Our findings confirm that ORCs are a rare and heterogenous population of radio sources. We confirm the association with large ellipticals in most cases, and we note a relation between the ORCs' physical sizes and their integrated spectral indexes, with small ORCs avoiding steep spectra.}


\maketitle

\nolinenumbers


\begin{table*}[!htb]
\centering
\caption{Summary of ORC properties.}\label{tab:sample}
\begin{threeparttable}
\begin{tabular}{lcccccc l}
\hline
Name & RA [deg] & Dec [deg] & $z$ & $R_{\rm proper}$ [kpc] & Sp. Index & $L_{144}$ [W Hz$^{-1}$] & Class. \\
\hline
J0016+2426 & 4.118 & 24.436 & 0.375 & 100 & -0.9±0.1 & $(1.37\pm0.01)\times10^{26}$ & cand ORC \\
J0320+1610 & 50.039 & 16.176 & 0.345 & 45 & -0.72±0.01 & $(1.79\pm0.01)\times10^{26}$ & cand ORC \\
J0324+8322 & 51.210 & 83.367 & 0.365 & 174 & -0.97±0.05 & $(1.16\pm0.01)\times10^{26}$ & cand ORC \\
J0801+5544 & 120.265 & 55.740 & 0.770 & 207 & -1.4±0.1 & $(4.44\pm0.02)\times10^{26}$ & cand ORC \\
J0823+6216 & 125.924 & 62.273 & 0.039 & 44 & -- & $(2.20\pm0.05)\times10^{23}$ & ORC \\
J0825+6132 & 126.260 & 61.544 & 0.800 & 120 / 152 & -1.5±0.3 & $(2.94\pm0.18)\times10^{25}$ & DHORC \\
J0847+7022 & 131.784 & 70.369 & 0.405 & 308 / 184 & -1.3±0.3 & $(3.65\pm0.08)\times10^{25}$ & DHORC \\
J0940+6028 & 145.213 & 60.480 & 0.533 & 161 & -1.18±0.03 & $(1.87\pm0.01)\times10^{26}$ & cand ORC \\
J1134+6642 & 173.740 & 66.716 & 0.400 & 339 & -0.5±0.3$^a$ & $(2.47\pm0.09)\times10^{25}$ & ORC \\
J1150+2449 & 177.585 & 24.821 & 0.670 & 127 & -- & $(1.57\pm0.11)\times10^{25}$ & ORC \\
J1222+6436 & 185.544 & 64.611 & 0.245 & 112 & -- & $(1.02\pm0.09)\times10^{24}$ & ORC \\
J1313+5003 & 198.446 & 50.055 & 0.860 & 136 / 148 & -1.3±0.2 & $(1.53\pm0.02)\times10^{26}$ & DHORC \\
J1436+4832 & 219.081 & 48.536 & 0.191 & 107 & -1.14±0.03 & $(3.66\pm0.03)\times10^{25}$ & cand ORC \\
J1458+4534 & 224.557 & 45.568 & -- & -- & -1.73±0.06 & -- & cand ORC \\
J1559+2734 & 239.971 & 27.577 & -- & -- & -0.77±0.07 & -- & cand ORC \\
J1608+6123 & 242.141 & 61.399 & 0.340 & 96 & -0.92±0.08 & $(2.11\pm0.04)\times10^{25}$ & cand ORC \\
J1633+1441 & 248.434 & 14.699 & 0.128 & 208 & -- & $(3.47\pm0.15)\times10^{24}$ & ORC \\
J1555+2726 (ORC4) & 238.853 & 27.443 & 0.451 & 184 / 186 & -1.3±0.2 & $(4.33\pm0.06)\times10^{25}$ & DHORC \\
\hline
\end{tabular}
\begin{tablenotes}
\item[a] The integrated spectral index is dominated by the central source. Removing its contribution, we estimate $\alpha=-1.2\pm0.3$. See source description in the text.
\end{tablenotes}
\end{threeparttable}
\end{table*}

\section{Introduction}
\label{sec:introduction}

The first odd radio circle (ORC) was noticed in 2013 by Intema in archival observations of the cluster Abell 2142 \citep{Venturi2017}. The recognition of the source peculiarity happened some years later thanks to the seminal paper of \citet{Norris2021a}. The defining characteristic of the sources is a ring (or edge-brightened disk) of radio emission that does not currently have any counterpart at other wavelengths and is associated with a galaxy near its center \citep{Norris2025}. The presence of a massive ($> 10^{11} \Msun$) elliptical galaxy close to the rings' centers is a property of ORCs \citep[e.g.][]{Rupke2024}. The host galaxies of ORCs also tend either to lie in overdense regions or have a close companion \citep{Norris2021b}.

\cite{Norris2021a} identified two ORCs using the Australian Square Kilometre Array Pathfinder (ASKAP) radio continuum data from the Evolutionary Map of the Universe (EMU) survey at $800-1088$~MHz \citep{Norris2011}. A further ORC was then reported by \cite{Koribalski2021}, again when inspecting ASKAP data, and two more were found via MeerKAT observations \citep{Norris2025, Koribalski2024a}; this brought the total count to five single ORCs. A number of other ORCs were also claimed in \cite{Gupta2025}.

Recently, double-headed ORCs (DHORCs), i.e., two nearby or superimposed ORCs related to the same formation mechanism and associated with one galaxy, were detected. The most prominent case is RAD~J131346.9+500320, identified in LOFAR Two Metre Sky Survey \citep[LoTSS;][]{Shimwell2022} data: it consists of two intersecting ring-like radio structures centered on a host galaxy at redshift $z\sim 0.9$ \citep{Hota2025}. This double ORC presents several common ORC traits, including large diameters, steep radio spectra, and sharp intensity gradients. Each ring is $\sim 300$ kpc in diameter, with diffuse emission extending over 800 kpc; the morphology suggests a shell originated by a shock-front. It has a steep radio spectrum with a spectral index\footnote{Defined as $S_{\nu} \propto \nu^{\alpha}$, with $S_{\nu}$ being the flux density and $\nu$ the observing frequency.} of $\sim -1.2$. Other examples of DHORCs are ORC J0356-4216 \citep{Taziaux2025} and ORC J1841-6547 \citep{Koribalski2025}. For the former a polarimetric analysis was carried out, finding that it is 20-30\% polarized, with a magnetic-field strength estimated at $1.6-1.8$~$\mu$G and magnetic-field orientations tangential to the rings. These peculiar sources suggest that ORCs may be a heterogeneous class of extragalactic radio sources.

While the origin of ORCs is still unclear, \citet{Norris2022} and \citet{Koribalski2021} already proposed some models that reproduce (at least some of) the observational properties. However, the physical nature and formation of ORCs remain subjects of active research and debate. Multiple scenarios have been investigated with simulations:
\begin{itemize}
    \item Spherical shock waves produced by energetic, transient events in galaxies—such as mergers of supermassive black holes \citep[e.g.][]{Yamasaki2024, Norris2021a}, intense starburst-driven outflows \citep[e.g.][]{Coil2024, Yamasaki2024}, or galaxy mergers \citep[e.g.][]{2018ApJ...857...50Y, Dolag2023a, Yamasaki2024, Koribalski2024, Ivleva2026}. These shocks may accelerate relativistic particles in the circumgalactic medium or the intergalactic medium, producing the observed synchrotron emission. Among the cited scenarios, the galaxy merger shock model was explored in more detail. The total radio power predicted by this model lies at around $10^{21}$ W/Hz at 150 MHz, and it is at least three orders of magnitude lower than powers inferred from observations \citep[see Fig.~7 in][]{Ivleva2026}. This mismatch may be caused by i) missing magnetic-field amplification mechanisms at the shock and/or ii) missing fossil CR populations that can be reaccelerated and contribute to the ORC radio power.
    \item Remnants of galactic outflows (or outflowing galactic radio-emitting structures; OGREs), in which energetic ejections from galaxies' active nuclei drive shock waves that accelerate cosmic rays \citep{Fujita2024}.
    \item Reactivated radio lobes from previous AGN activity, reignited by shocks associated with mergers of galaxy groups or clusters \citep{Shabala2024, Wang2026}. While the CR population injected by AGN fades quickly, in case of reacceleration by a shock travelling through a preexisting AGN bubble, the radio morphology matches well with observed ORCs.
    \item Synchrotron emission from cosmological virial shocks, where relativistic electrons are accelerated within galaxy halos \citep{Yamasaki2024}.
\end{itemize}

Several of the known ORCs have a rather steep spectral index (usually $\lesssim -1$), making low-frequency radio surveys the perfect hunting ground for these elusive sources. The typical apparent size of ORCs is around 1\arcmin{} (usually corresponding to a few hundred kiloparsecs, depending on the source redshift), requiring an angular resolution of a few arcseconds or better in order to classify them morphologically. Finally, their rarity requires a search in large-area surveys. LoTSS fulfils these requirements with a central frequency of 144 MHz, a resolution of 5\arcsec{}, and a large footprint covering the entire extragalactic northern hemisphere. In this study we used this survey to search for new ORCs.

The paper is organized as follows. In Sect.~\ref{sec:observations} we describe the dataset and the selection of the sources. In Sect.~\ref{sec:single} we go through all the ORCs and candidate ORCs, describing their characteristics. Our discussion and conclusions are presented in Sects.~\ref{sec:discussion} and~\ref{sec:conclusions}, respectively. Throughout this paper we assume a $\Lambda$CDM cosmological model from \citet{PlanckCollaboration2020} with $H_0 = 67.4~\mathrm{km/s/Mpc}$, $\Omega_M = 0.315$. For the convention of the spectral index, $\alpha$, we used $S_\nu \propto \nu^{\alpha}$. 

\begin{table*}[tb]
\centering
\caption{Host galaxies and best redshift estimation from spectroscopic or combined photometric observations.}\label{tab:opt}
\begin{threeparttable}
\begin{tabular}{l c l l}
\hline
Name & $z$ & Host galaxy & $z$ from literature\\
\hline
J0016+2426 & 0.37534 & SDSS J001628.29+242610.3 $^{a}$ & Spectroscopic redshift$^{6}$ \\
J0320+1610 & 0.345 & PSO J050.0394+16.1762 & 0.340$^{4}$, 0.352$^{7}$ \\ & & & 0.341$^{10}$, 0.50$^{15}$ \\
J0324+8322 & 0.3655 & PSO J051.2103+83.3666 $^{a}$ & 0.375$^{7}$, 0.356$^{10}$ \\
J0801+5544 & 0.77 & SDSS J080103.48+554424.8 $^{a}$ & 0.800$^{5}$, 0.6924$^{3}$, -$^{4}$ \\ & & & 0.846$^{7}$, 0.768$^{9}$, 0.746$^{11}$, 0.770$^{8}$ \\
J0823+6216 & 0.03858 & 2MASX J08234191+6216214 & Spectroscopic redshift$^{1}$ \\
J0825+6132 & 0.8 & DESI J126.2599+61.5444 $^{a}$ & 0.785$^{7}$, 0.816$^{10}$, 0.836$^{9}$ \\ & & & 0.834$^{11}$, 0.781$^{16}$, 0.830$^{8}$ \\
J0847+7022 & 0.405 & DESI J131.7838+70.3686 & 0.344$^{4}$, 0.433$^{7}$ \\ & & & 0.438$^{10}$, 0.50$^{15}$, 0.404$^{8}$ \\
J0940+6028 & 0.53327 & SDSS J094051.23+602846.7 $^{a}$ & Spectroscopic redshift$^{6}$ \\
J1134+6642 & 0.4 & SDSS J113457.67+664256.7 & 0.466$^{5}$, 0.3765$^{3}$, 0.3385$^{4}$, 0.400$^{7}$ \\ & & & 0.407$^{10}$, 0.386$^{9}$, 0.473$^{11}$, 0.477$^{13}$, 0.3931$^{8}$ \\
J1150+2449 & 0.67 & SDSS J115020.44+244913.8 $^{a}$ & 0.737$^{5}$, 0.711$^{7}$, 0.637$^{10}$ \\ & & & 0.664$^{9}$, 0.663$^{11}$, 0.635$^{13}$, 0.683$^{16}$, 0.655$^{8}$ \\
J1222+6436 & 0.2449 & SDSS J122210.59+643638.9 $^{a}$ & Spectroscopic redshift$^{12}$ \\
J1313+5003 & 0.86 & SDSS J131346.92+500319.3 & 0.706$^{5}$, 0.813$^{7}$ \\ & & & 0.908$^{10}$, 0.937$^{9}$, 0.822$^{11}$, 0.923$^{16}$, 0.818$^{8}$ \\
J1436+4832 & 0.19115 & SDSS J143619.42+483210.5 & Spectroscopic redshift$^{6}$ \\
J1458+4534 & -- & -- & -- \\
J1559+2734 & -- & -- & -- \\
J1608+6123 & 0.34 & SDSS J160833.95+612356.8 $^{a}$ & 0.322$^{5}$, 0.3292$^{3}$, 0.319$^{4}$, 0.340$^{7}$, 0.331$^{10}$ \\ & & & 0.328$^{9}$, 0.341$^{11}$, 0.3545$^{2}$, 0.350$^{16}$, 0.343$^{8}$ \\
J1633+1441 & 0.1281 & SDSS J163344.16+144154.9 & Spectroscopic redshift$^{6}$ \\
J1555+2726 (ORC4) & 0.4512 & SDSS J155524.63+272634.3 & Spectroscopic redshift$^{14}$ \\
\hline
\end{tabular}
\begin{tablenotes}
\item \textsuperscript{a} Uncertain association.
    \textsuperscript{1} \citet{1999PASP..111..438F}; \textsuperscript{2} \citet{2011ApJ...736...21S}; \textsuperscript{3} \citet{2014A&A...568A.126B}; \textsuperscript{4} \citet{2016ApJS..225....5B}; \textsuperscript{5} \citet{2016MNRAS.460.1371B}; \textsuperscript{6} \citet{2020ApJS..249....3A}; \textsuperscript{7} \citet{2021MNRAS.500.1633B}; \textsuperscript{8} \citet{2021MNRAS.501.3309Z}; \textsuperscript{9} \citet{2022MNRAS.512.3662D}; \textsuperscript{10} \citet{2022MNRAS.515.4711B}; \textsuperscript{11} \citet{2022RAA....22f5001Z}; \textsuperscript{12} \citet{2023A&A...674A..31D}; \textsuperscript{13} \citet{2024ApJS..272...39W}; \textsuperscript{14} \citet{Coil2024}; \textsuperscript{15} \citet{2024OJAp....7E...6F}; \textsuperscript{16} \citet{2025MNRAS.536.2260Z}.
\end{tablenotes}
\end{threeparttable}
\end{table*}

\begin{table}[tb]
\centering
\caption{Measurements of flux densities and relative errors at 54, 144, and 1400 MHz.}\label{tab:fluxes}
\begin{threeparttable}
\begin{tabular}{lccc}
\hline
Name & $S_{54}$ [mJy] & $S_{144}$ [mJy] & $S_{1400}$ [mJy]\\
\hline
J0016+2426 & 594 $\pm$ 14 & 265 $\pm$ 1 & 25 $\pm$ 1 \\
J0320+1610 & -- & 421 $\pm$ 1 & 81 $\pm$ 2 \\
J0324+8322 & 655 $\pm$ 15 & 237 $\pm$ 1 & 28 $\pm$ 1 \\
J0801+5544 & 667 $\pm$ 14 & 153 $\pm$ 1 & 10 $\pm$ 1 \\
J0823+6216 & -- & 59 $\pm$ 1 & -- \\
J0825+6132 & 41 $\pm$ 11 & 9.2 $\pm$ 0.6 & -- \\
J0847+7022 & 242 $\pm$ 5 & 59 $\pm$ 1 & 12 $\pm$ 2 \\
J0940+6028 & 506 $\pm$ 16 & 156.9 $\pm$ 0.9 & 11 $\pm$ 1 \\
J1134+6642 & 108 $\pm$ 6 & 41 $\pm$ 2 & 21 $\pm$ 2 \\
J1150+2449 & -- & 7.6 $\pm$ 0.5 & -- \\
J1222+6436 & -- & 5.2 $\pm$ 0.4 & -- \\
J1313+5003 & 172 $\pm$ 14 & 40.3 $\pm$ 0.6 & 4 $\pm$ 1 \\
J1436+4832 & 1000 $\pm$ 33 & 330 $\pm$ 3 & 24 $\pm$ 3 \\
J1458+4534 & 108 $\pm$ 4 & 20 $\pm$ 1 & -- \\
J1559+2734 & -- & 36.5 $\pm$ 0.7 & 6 $\pm$ 1 \\
J1608+6123 & 118 $\pm$ 16 & 51.4 $\pm$ 0.9 & 6 $\pm$ 1 \\
J1633+1441 & -- & 75 $\pm$ 3 & -- \\
J1555+2726 & 220 $\pm$ 14 & 54.3 $\pm$ 0.7 & 6 $\pm$ 1 \\

\hline
\end{tabular}
\end{threeparttable}
\end{table}

\section{Observations}
\label{sec:observations}

We explored the third release of the LOFAR Two Metre Sky Survey at 144 MHz \citep[LoTSS DR3;][]{Shimwell2026} searching for candidate ORCs. Given the few examples of known sources of this type, the possibility to train an automatic algorithm to detect them is limited. Attempts to isolate sources with peculiar morphology still require a further visual inspection \citep{Gupta2022}. We therefore decided to isolate a manageable number of sources from the survey catalog and inspect them by eye. The cut on the catalog was made by setting a total flux/peak flux ratio of $> 10$ to isolate extended sources; then, we limited the samples to sources of at least 40\arcsec{} in size and in regions with a low local rms (less than 0.1~\mjybeam). A final cut on Galactic latitude ($|b|>20\deg$) was applied to remove possible contamination from supernova remnants that may also exhibit shell-like morphologies. We note that a number of good candidates could be missed because the source finder may fragment the extended emission.

As discussed by other authors \citep[e.g.][]{Hota2025}, several sources can be easily misclassified as ORCs. these include lobes of radio galaxies, wide-angle tail (WAT) radio galaxies, dying radio galaxies, and nearby star-forming galaxies. In our classification we searched for round objects (composed of one or two circles) with edge-brightened surface brightness and with or without a central bright radio source.

From the LoTSS survey we discarded a large number of diffuse radio-emitting systems surrounding distant galaxies, most of the time in relatively modest overdensities such as galaxy groups. These sources possibly represent early ORC evolutionary stages or distinct evolutionary pathways, which are known as GLAREs \citep[Galaxies with Large-scale Ambient Radio Emission; e.g.][]{Gupta2022}.

To further optimize the image quality of some of the selected sources lying in particularly difficult LoTSS fields, we applied the same extraction process first described in \citet{vanWeeren2020}. In short, a relatively small region, typically $15-20$ arcmin, depending on the local integrated flux density, is selected around the target. Then, all sources outside this region are subtracted from the visibilities. After phase-shifting to the target position, averaging in time and frequency, and primary beam correction, additional self-calibration cycles are applied to the extracted field to further reduce the noise and improve the image quality. This process was applied to LoTSS data of J0016+2426, J0847+7022, and J0940+6028.

In order to obtain spectral information we retrieved data from the NRAO VLA Sky survey at 1400 MHz \citep[NVSS;][]{Condon1998} and from the LOFAR LBA Sky Survey at 54 MHz \citep[LoLSS;][]{deGasperin2021, DeGasperin2023} second data release (De Gasperin et al. in prep.). Finally, we made use of data from the Faint Images of the Radio Sky at Twenty Centimeters at 1400 MHz \citep[FIRST;][]{Becker1995} and the Karl G. Jansky Very Large Array Sky Survey \citep[VLASS;][]{Lacy2020} surveys to search for and isolate compact radio emission in the ORC region. From FIRST and VLASS we do not expect any emission detected on scales larger than 1\arcmin{} and 30\arcsec{}, respectively.

\section{Single source analysis}
\label{sec:single}

We describe the main characteristics of each source identified as an ORC, DHORC, or candidate ORC. The defining characteristic of ORCs is a prominent circular/elliptical shape with edge brightening.  When an optical counterpart could not be identified, or the circle was only partial or dominated by a strong localized region of emission, we deemed the source a "candidate" ORC. A summary of the source properties is listed in Table~\ref{tab:sample}, while information on the optical counterpart and radio flux densities is listed in Table~\ref{tab:opt} and Table~\ref{tab:fluxes}, respectively. Flux density errors are statistical based on the local rms noise multiplied by the square root of the number of beams covering the source extension.

\subsection{ORCs and DHORCs}

For each of the five ORCs and four DHORCs listed in this section we provide a short description, images, and brightness profiles.

\begin{figure}[htb!]
    \centering
    \includegraphics[width=.5\textwidth]{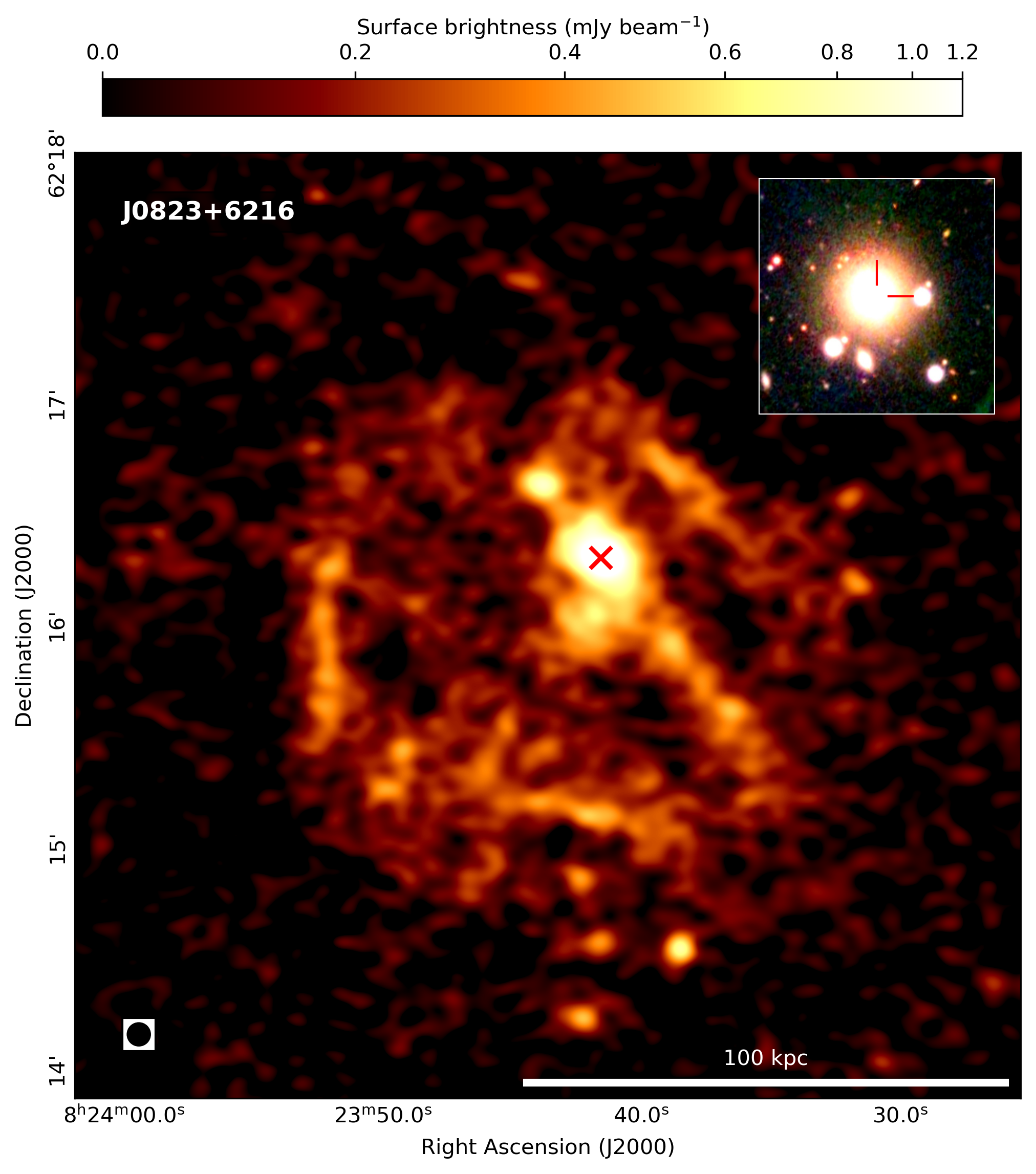}
    \includegraphics[width=.5\textwidth]{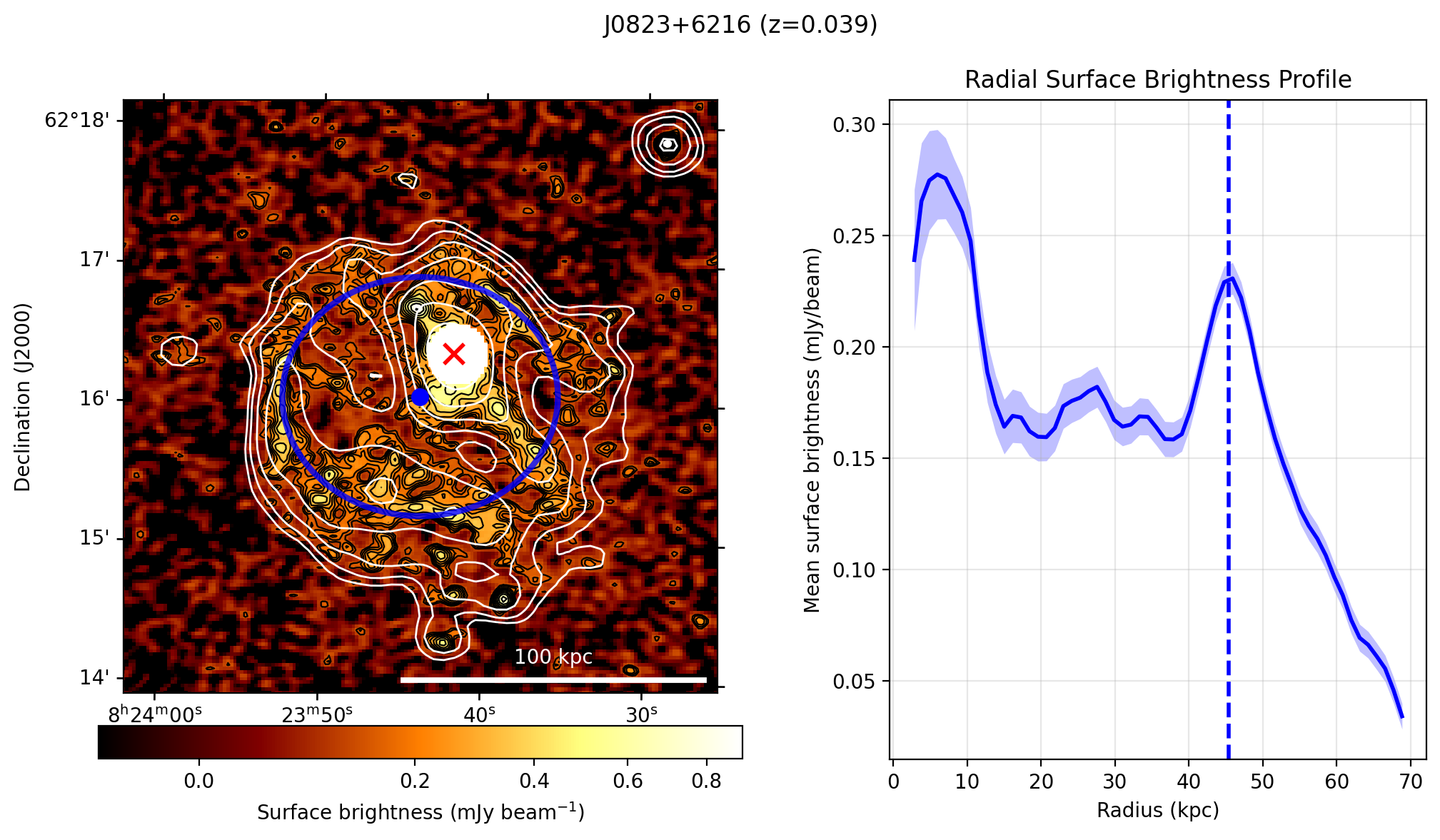}
    \caption{Top panel: Image at 144 MHz ORC J0823+6216. The resolution is 5\arcsec, with the beam shape visible in the bottom left of the image. The insert shows the optical field from PanSTARR (DESI for all other images) with the identified optical counterpart. The red "x" in the radio image marks the position of the optical counterpart. Bottom panels: Images at 144 MHz and brightness profiles. Contours start from $3\sigma$ and end at the maximum value of the color bars with a square root scaling. White contours trace the low-resolution (20\arcsec) 144 MHz brightness. Pixels are blanked (white) where a compact point source is identified within the ORC's confines. The green/blue dots identify the center of the green/blue ellipses that trace the ORC's shape. The right part of the panel shows the profile along elliptical annuli with the vertical dashed lines drawn at the length of the ellipse's semi-major axis. The correspondence with a peak proves the enhancement of brightness along the ellipses line.}
    \label{fig:J0823+6216}
\end{figure}

\paragraph{ORC J0823+6216 (Fig.~\ref{fig:J0823+6216})} The source shows an elliptical pattern that is rather well defined for the entire edge. The brightness excess in the ORC region is clearly visible in the radial profile. Another thin structure is visible within the circle's edge. It has a straight morphology and overlaps with a bright, fairly compact source, which might indicate the presence of multiple circles seen edge-on and face-on. The compact emission lies within the ORC circle, but not at its centre and has no counterpart in FIRST or VLASS. The compact radio emission comes from the brightest and closest optical counterpart in the center of the ORC. The optical source is the very bright ($r=13.57$ mag) elliptical galaxy 2MASX J08234191+6216214 at $z=0.03858$. The source was not detected in other radio surveys. Compared to other known ORCs, the source has a relatively small radius of $\sim 44$~kpc.

\begin{figure}[htb!]
    \centering
    \includegraphics[width=.5\textwidth]{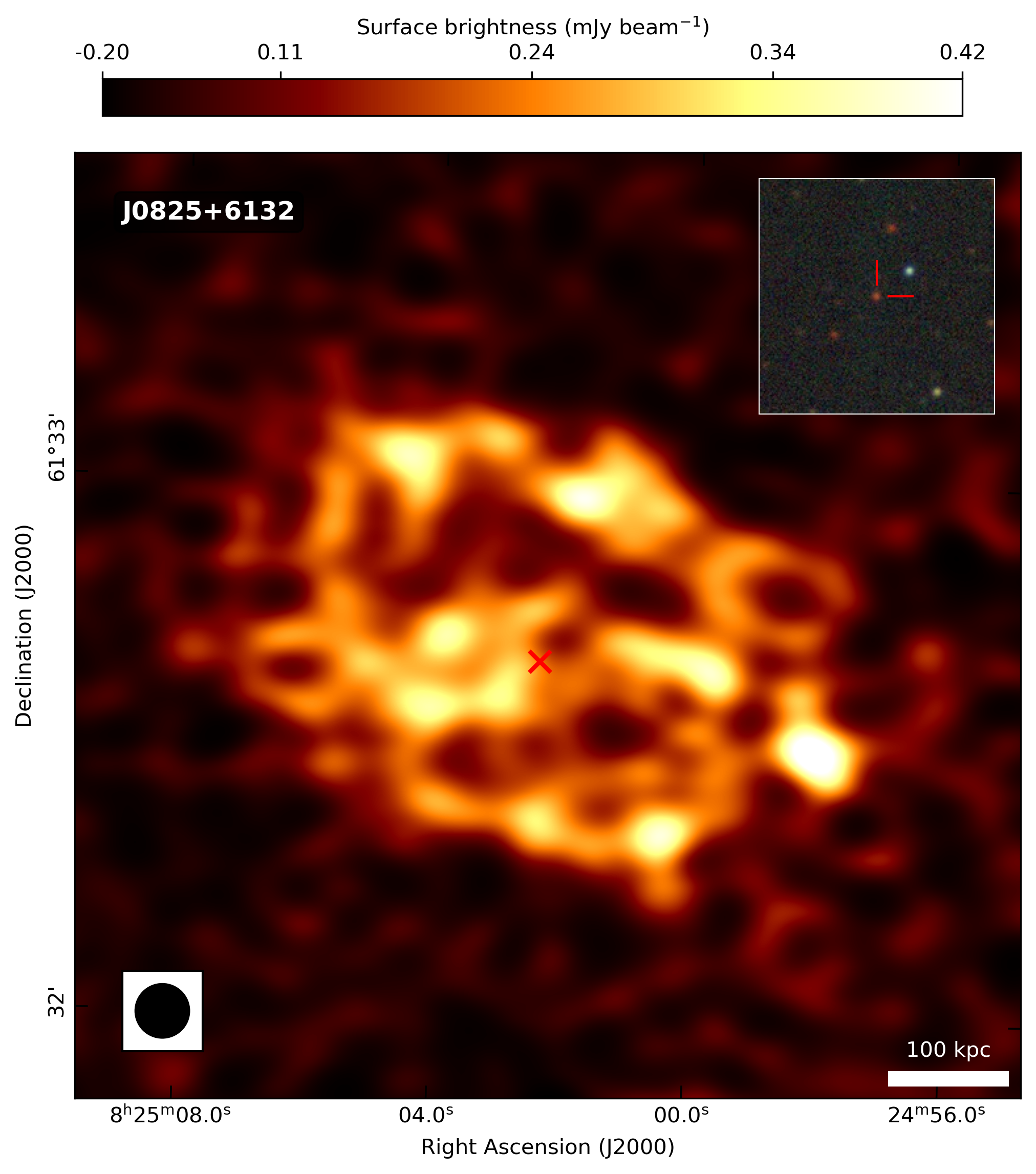}
    \includegraphics[width=.5\textwidth]{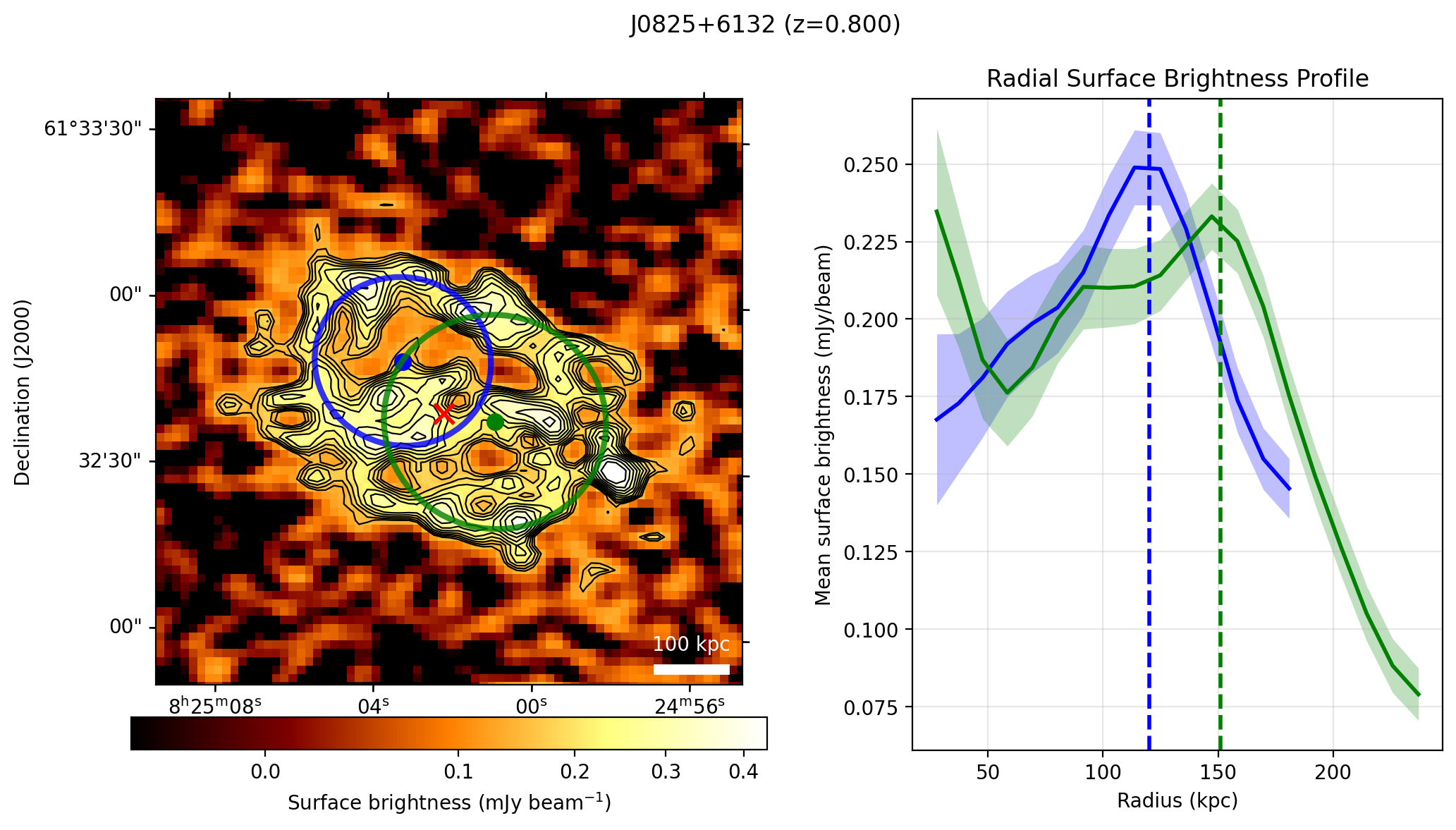}
    \caption{Same as Fig.~\ref{fig:J0823+6216}, but for ORC J0825+6132.}
    \label{fig:J0825+6132}
\end{figure}

\paragraph{DHORC J0825+6132 (Fig.~\ref{fig:J0825+6132})}
This source is a double-headed ORC with two circular components. We tentatively identified an optical counterpart when searching for galaxies in the central region of the ORC (DESI J126.2599+61.5444) at $z=0.8$. At this redshift the two circular components have radii of 124 and 155 kpc, respectively. No clear bright, compact radio source is present in the DHORC region. FIRST and VLASS do not show any compact emission. The source is not detected in the NVSS. The low-frequency spectral index between LoLSS and LoTSS provides a value of $\alpha = -1.5\pm0.3$. 

\begin{figure}[htb!]
    \centering
    \includegraphics[width=.5\textwidth]{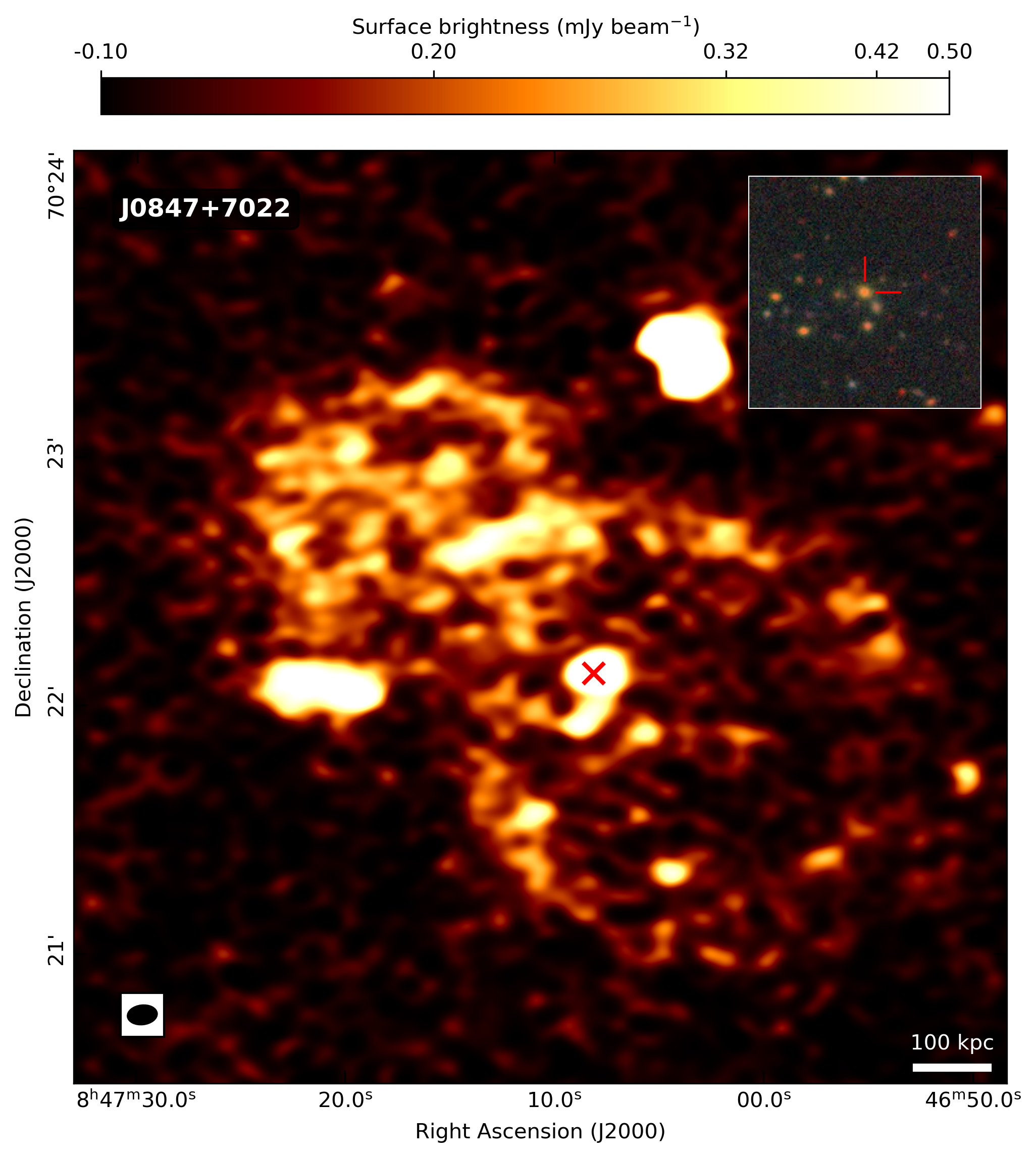}
    \includegraphics[width=.5\textwidth]{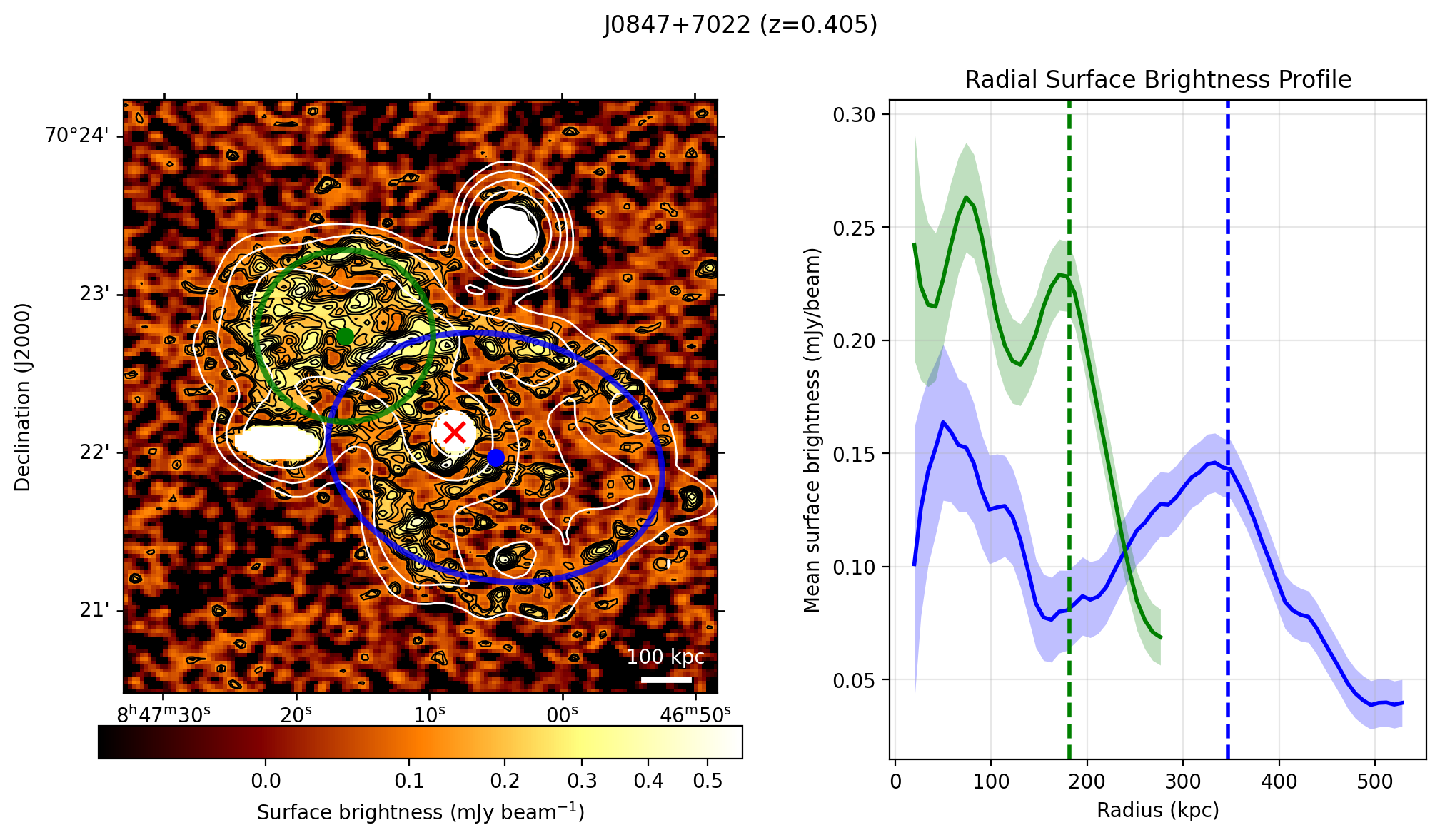}
    \caption{Same as Fig.~\ref{fig:J0823+6216} but for DHORC J0847+7022. Contours starts at $2\sigma$.}
    \label{fig:J0847+7022}
\end{figure}

\paragraph{DHORC J0847+7022 (Fig.~\ref{fig:J0847+7022})}
This is the second identified DHORC in the sample. The source has two thin structures. The first, in the NE direction, has a radio-filled circular shape, while the second, larger and fainter structure, in the SW direction, has an elliptical shape. The source has a bright compact radio emission close to its center that we identify with DESI J131.7838+70.3686 ($z=0.405$) as the optical counterpart.
At this redshift the radii of the components are 184 and 308 kpc, respectively. A search in NVSS resulted in a marginal detection, with a non-negligible fraction of the flux density coming from blended compact sources. The source is detected in LoLSS, but the emission is blended with a couple of compact sources in the vicinity. We attempted a low-frequency spectral index estimation using LoLSS and LoTSS, noting that the emission from the extended structure of the ORC is dominant, obtaining $\alpha=-1.4\pm0.2$.

\begin{figure}[htb!]
    \centering
    \includegraphics[width=.5\textwidth]{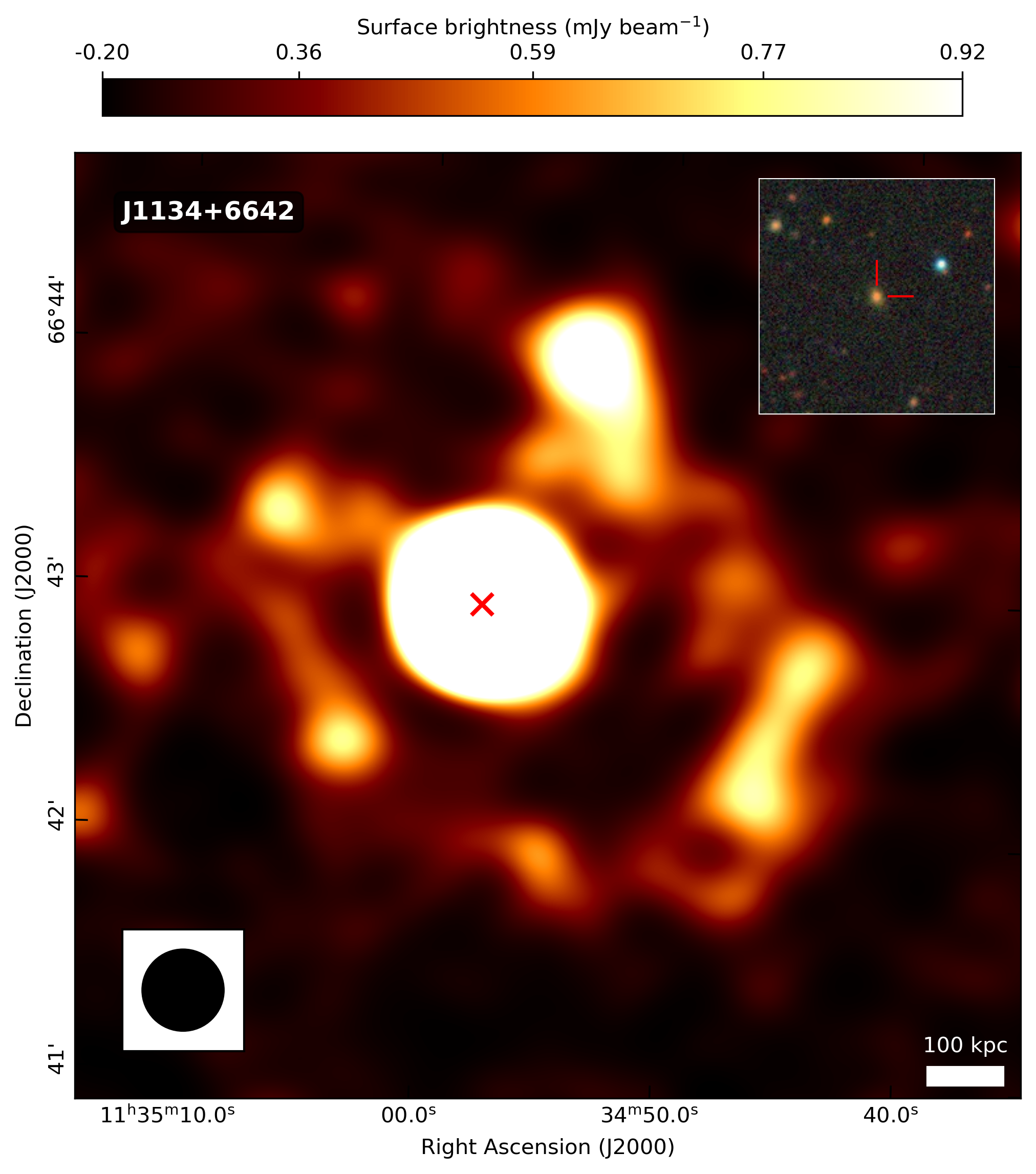}
    \includegraphics[width=.5\textwidth]{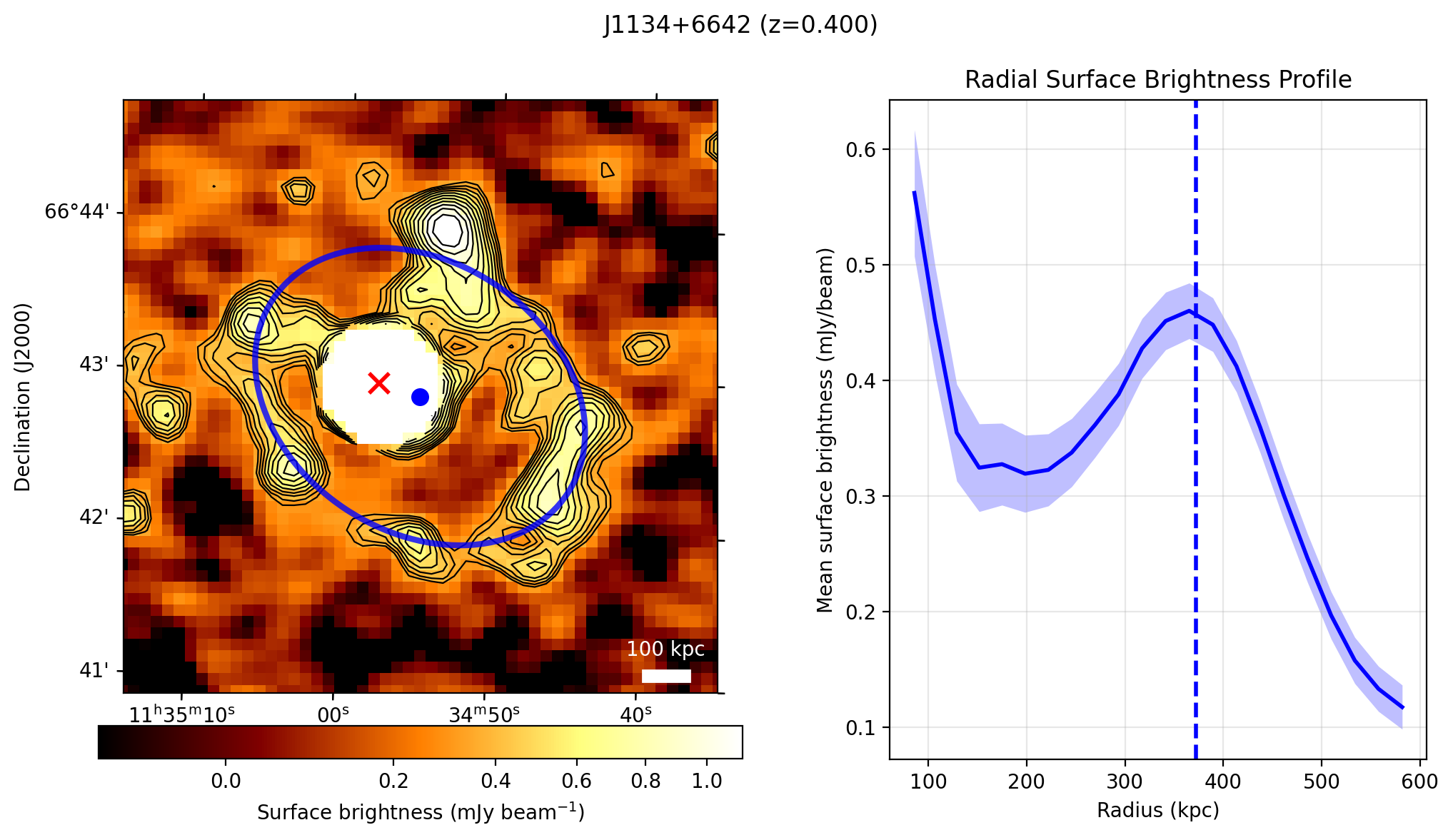}
    \caption{Same as Fig.~\ref{fig:J0823+6216}, but for ORC J1134+6642. The resolution is 20\arcsec.}
    \label{fig:J1134+6642}
\end{figure}

\paragraph{ORC J1134+6642 (Fig.~\ref{fig:J1134+6642})}
This source has the shape of a faint large ellipse with a comparatively bright central compact source displaced from the ellipse's center. The central compact source is associated with the optical counterpart SDSS J113457.67+664256.7 ($z=0.4$). The ellipse has a mean radius of 339 kpc. FIRST is not available in this part of the sky, and NVSS is strongly dominated by the central source. VLASS shows a point source with a flux density of $\sim11$~mJy. Using LoTSS and VLASS we derived a spectral index of the compact emission of $\alpha=-0.2$. The emission from the ring becomes more prominent at lower frequencies; hence, the concave spectrum is dominated by the compact emission at high frequencies and by the ring at lower frequencies. Using only LoLSS and LoTSS, we can obtain a spectral index that is less dominated by the central compact sources and treat it as a lower limit for the spectral index of the ring; we find $\alpha < -0.99$. If we estimate the flux density of the compact source at 144 and 54 MHz using the high-frequency spectral index and subtract it from LoTSS and LoLSS measurements, the estimation of the low-frequency spectral index of the extended emission becomes $\alpha = -1.2 \pm 0.3$.

\begin{figure}[htb!]
    \centering
    \includegraphics[width=.5\textwidth]{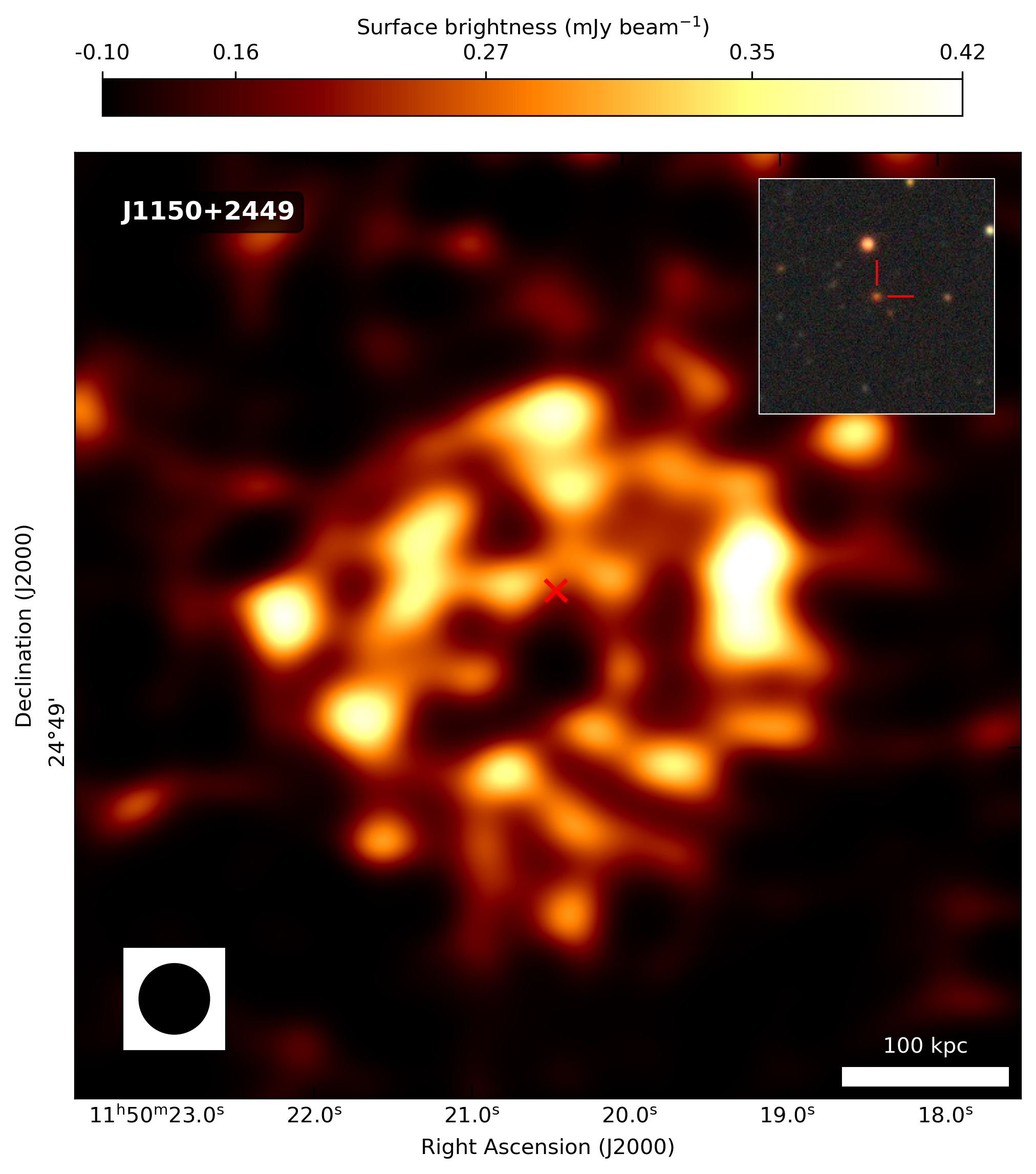}
    \includegraphics[width=.5\textwidth]{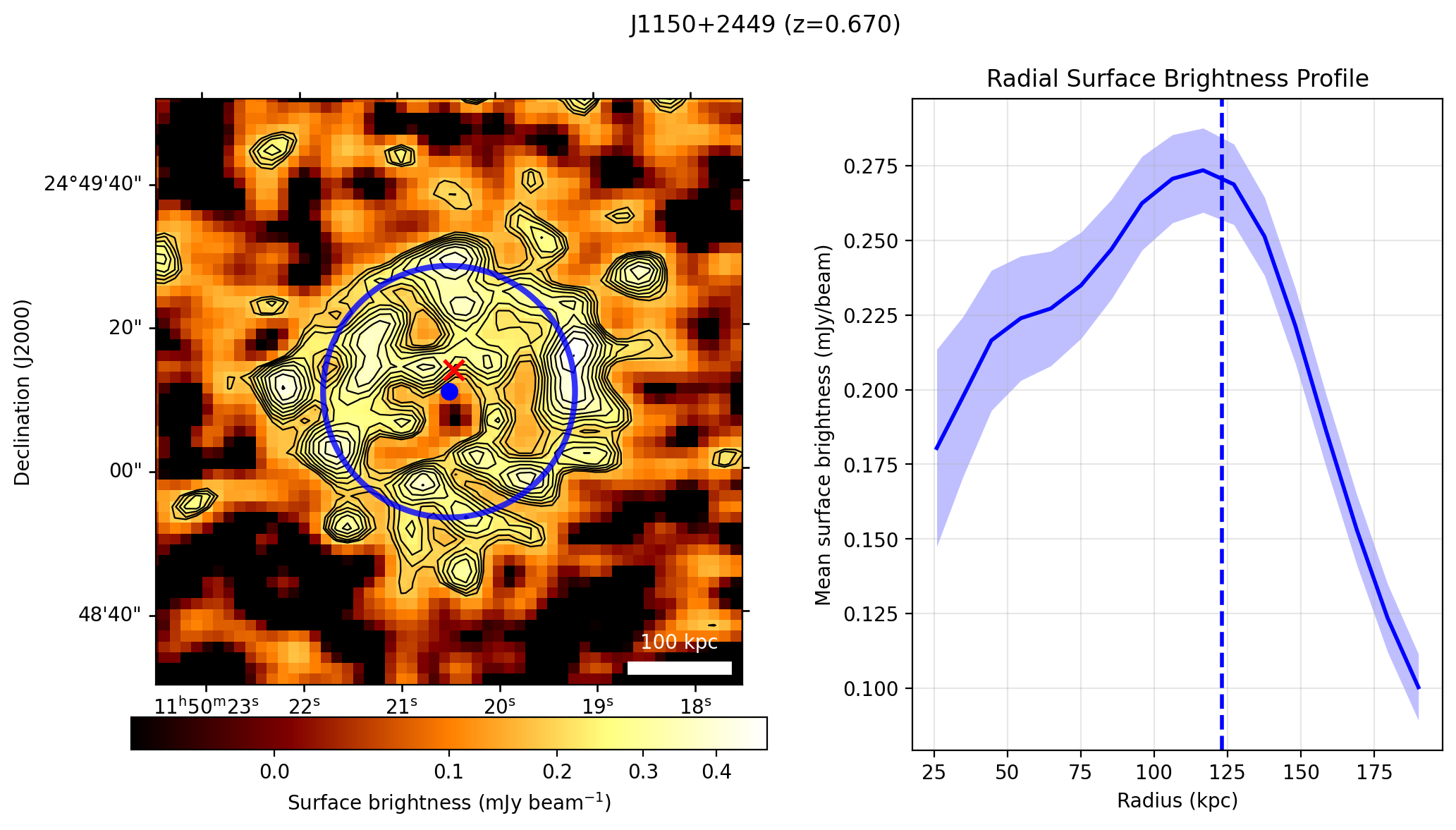}
    \caption{Same as Fig.~\ref{fig:J0823+6216}, but for ORC J1150+2449.}
    \label{fig:J1150+2449}
\end{figure}

\paragraph{ORC J1150+2449 (Fig.~\ref{fig:J1150+2449})}
This source is composed of one or possibly two rings in projection. No central, compact radio emission is detected in any survey, so the optical counterpart closest to the source center is SDSS J115020.44+244913.8 ($z=0.67$). At this redshift, the cleanest ring has a radius of 127 kpc.  No data are available at 54 MHz, and the source is not detected in NVSS. Hence, no information on the spectral index is available.

\begin{figure}[htb!]
    \centering
    \includegraphics[width=.5\textwidth]{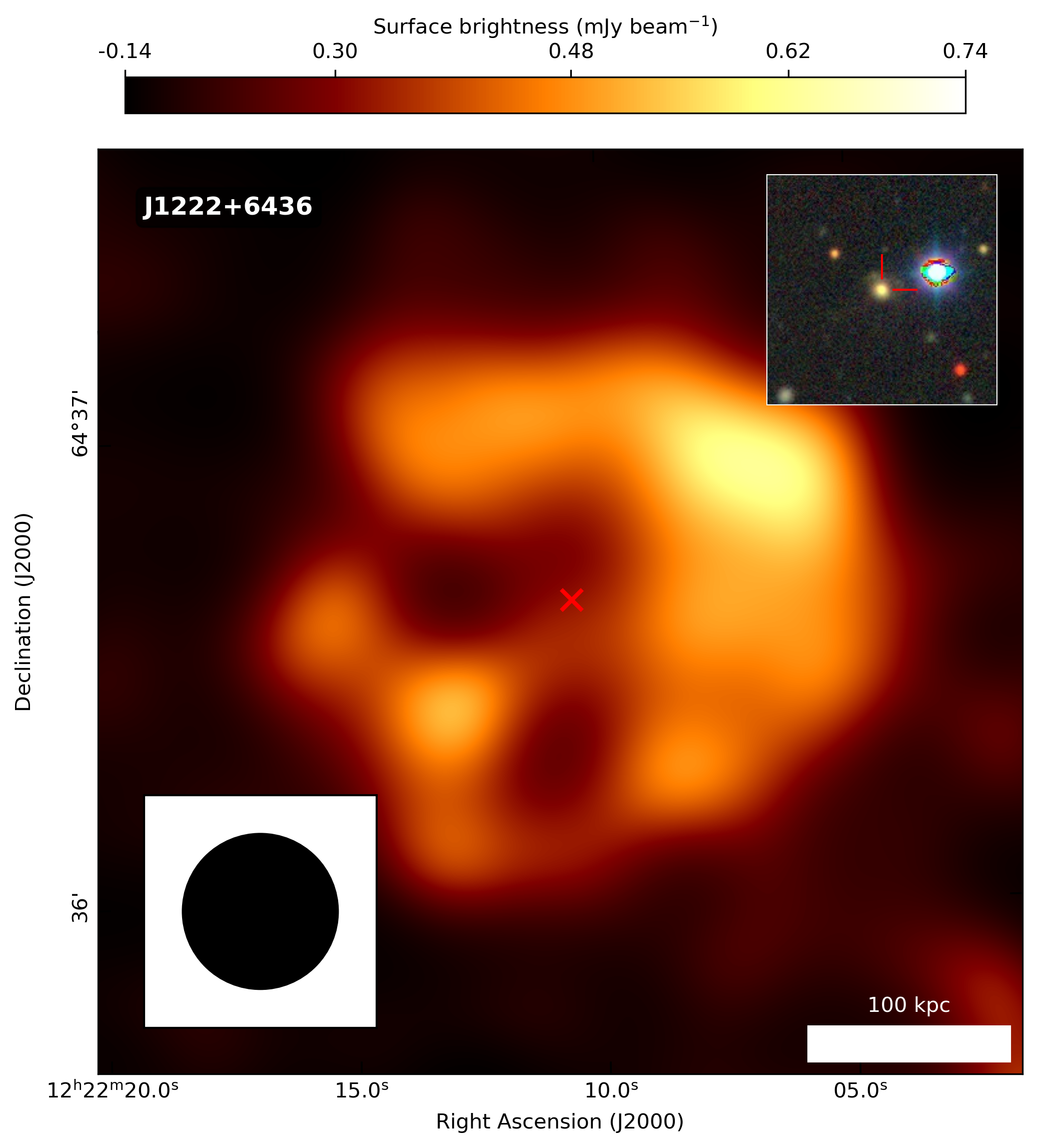}
    \includegraphics[width=.5\textwidth]{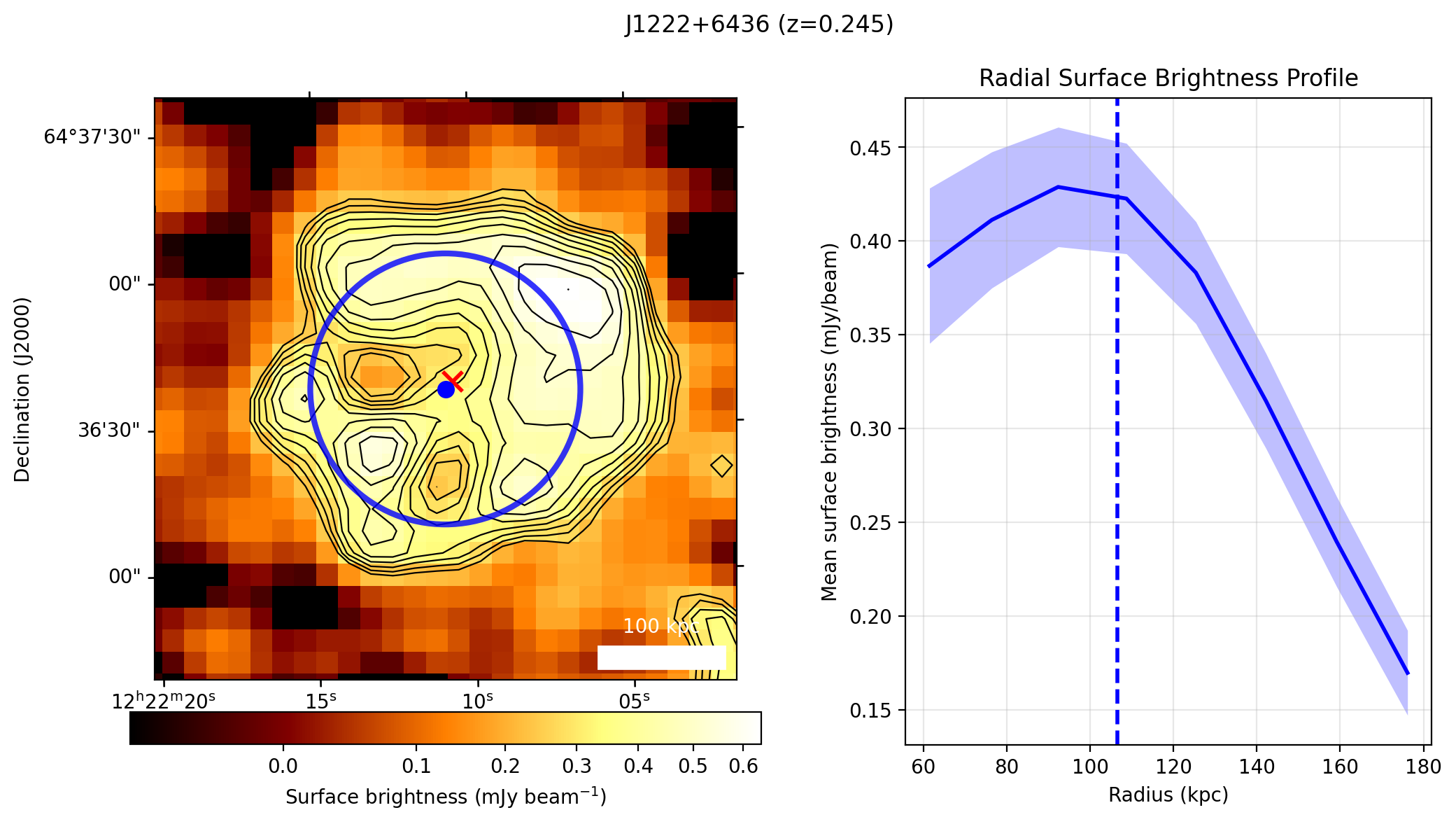}
    \caption{Same as Fig.~\ref{fig:J0823+6216}, but for ORC J1222+6436. The resolution is 20\arcsec.}
    \label{fig:J1222+6436}
\end{figure}

\paragraph{ORC J1222+6436 (Fig.~\ref{fig:J1222+6436})}
The source appears faint and diffuse at high resolution. To maximize the significance of the detection, we focussed the analysis on the low-resolution version of LoTSS at 20\arcsec. The image shows a clear ring with low brightness emission present also within the ring boundaries. We did not detect any compact source in any survey, and we selected the optical counterpart (SDSS J122210.59+643638.9, $z=0.2449$) as the closest galaxy to the ring center. No data are available at 54 MHz, and the source is not detected in NVSS; therefore, no information on the spectral index is available.

\begin{figure}[htb!]
    \centering
    \includegraphics[width=.5\textwidth]{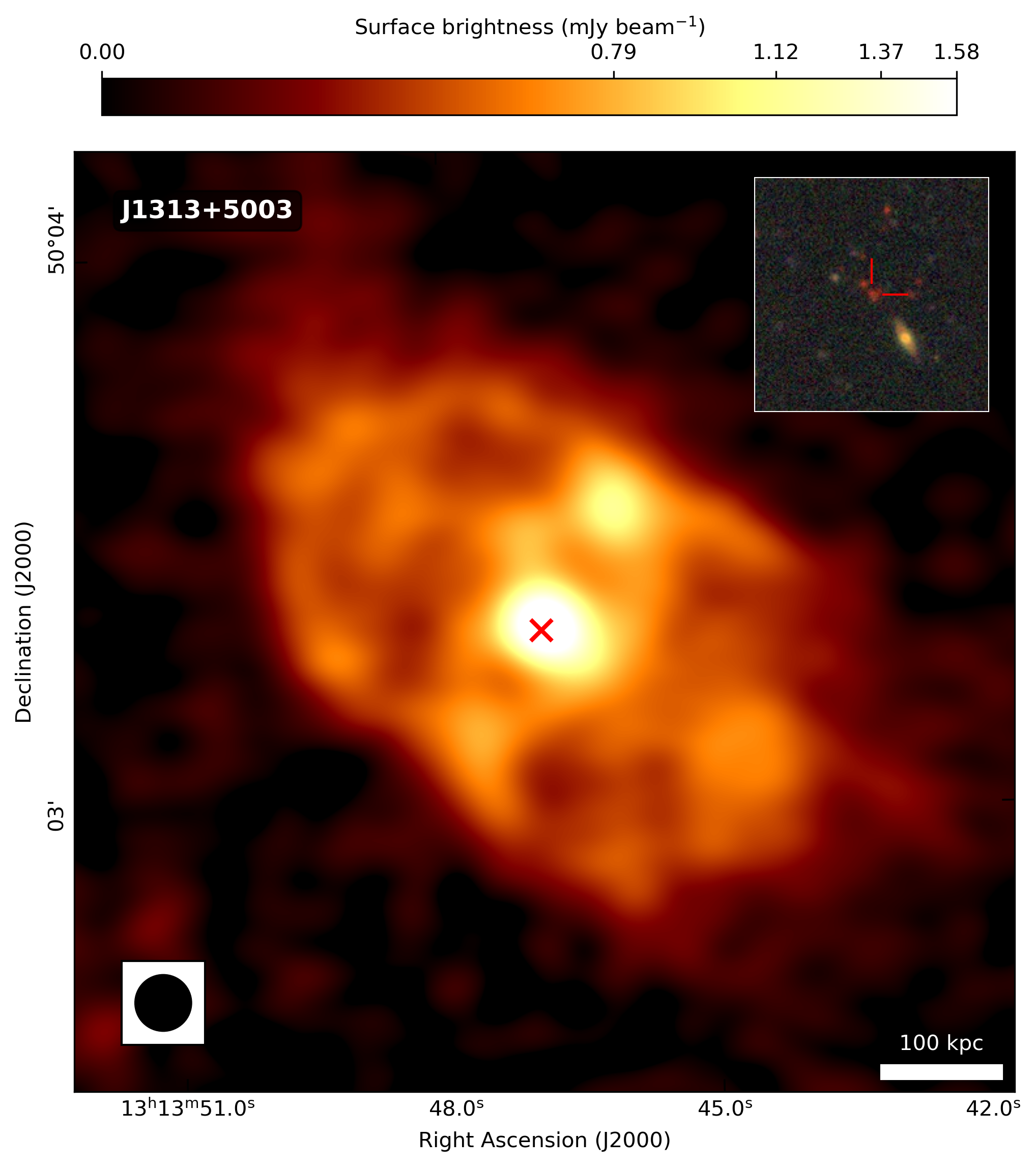}
    \includegraphics[width=.5\textwidth]{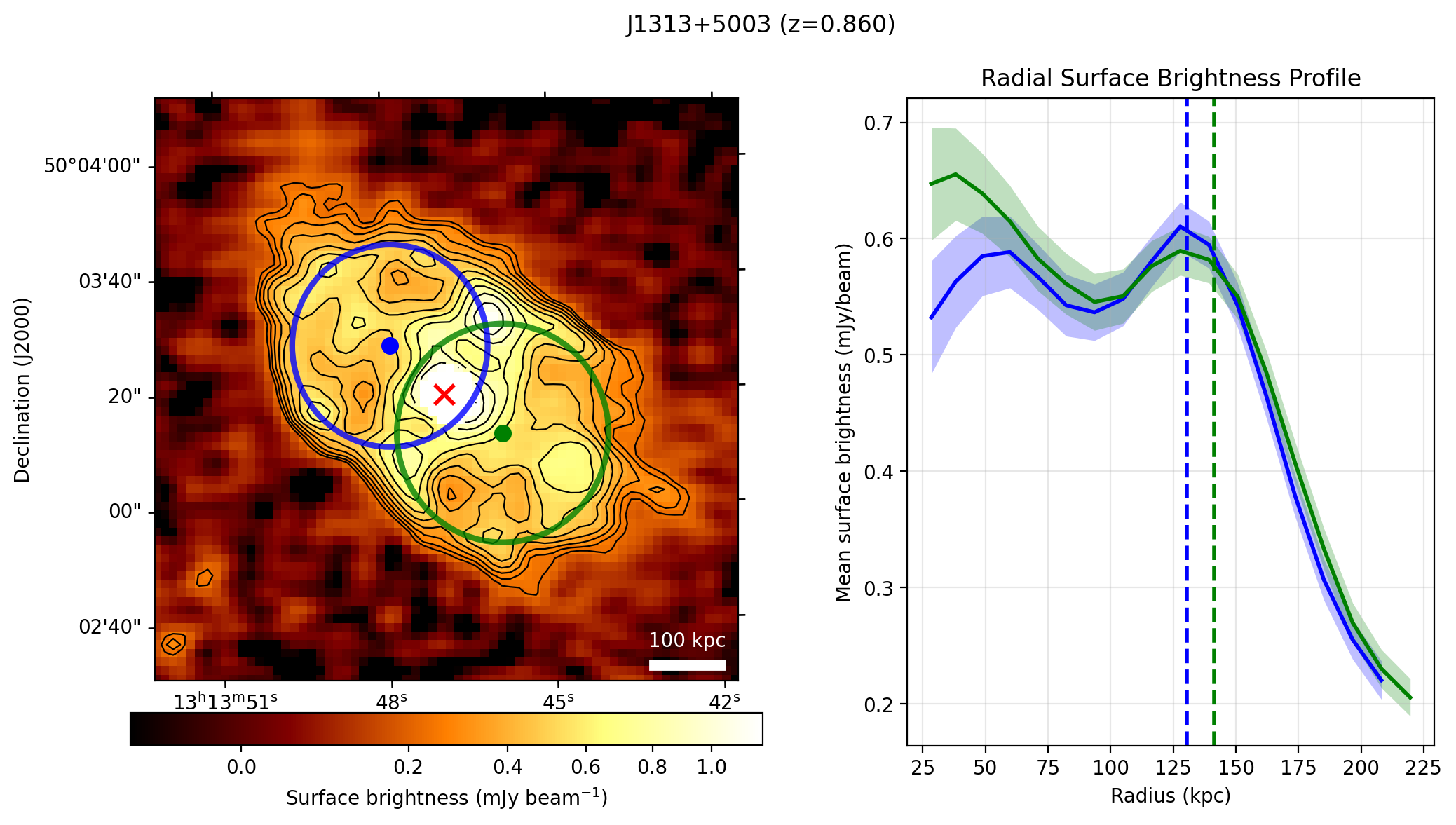}
    \caption{Same as Fig.~\ref{fig:J0823+6216}, but for DHORC J1313+5003.}
    \label{fig:J1313+5003}
\end{figure}

\paragraph{DHORC J1313+5003 (Fig.~\ref{fig:J1313+5003})}
This is the third DHORC of the sample, which was recently analysed by \cite{Hota2025}. The source has a central compact emission that we used to identify the optical counterpart with SDSS J131346.92+500319.3 ($z=0.86$). The source is elongated, and the morphology is well fit by two circular rings of similar radii ($\sim 140$ kpc). No detection is reported in FIRST or VLASS. The source is detected in LoLSS and NVSS. We used all three surveys to infer an integrated spectral index of $\alpha=-1.3\pm0.2$, which is in line with the value $\alpha=-1.2\pm0.1$ claimed by \citet{Hota2025}. However, using only the low-frequency data, the spectrum becomes slightly steeper ($\alpha=-1.5\pm0.1$), most likely due to the diminished contribution of the central compact source.

\begin{figure}[htb!]
    \centering
    \includegraphics[width=.5\textwidth]{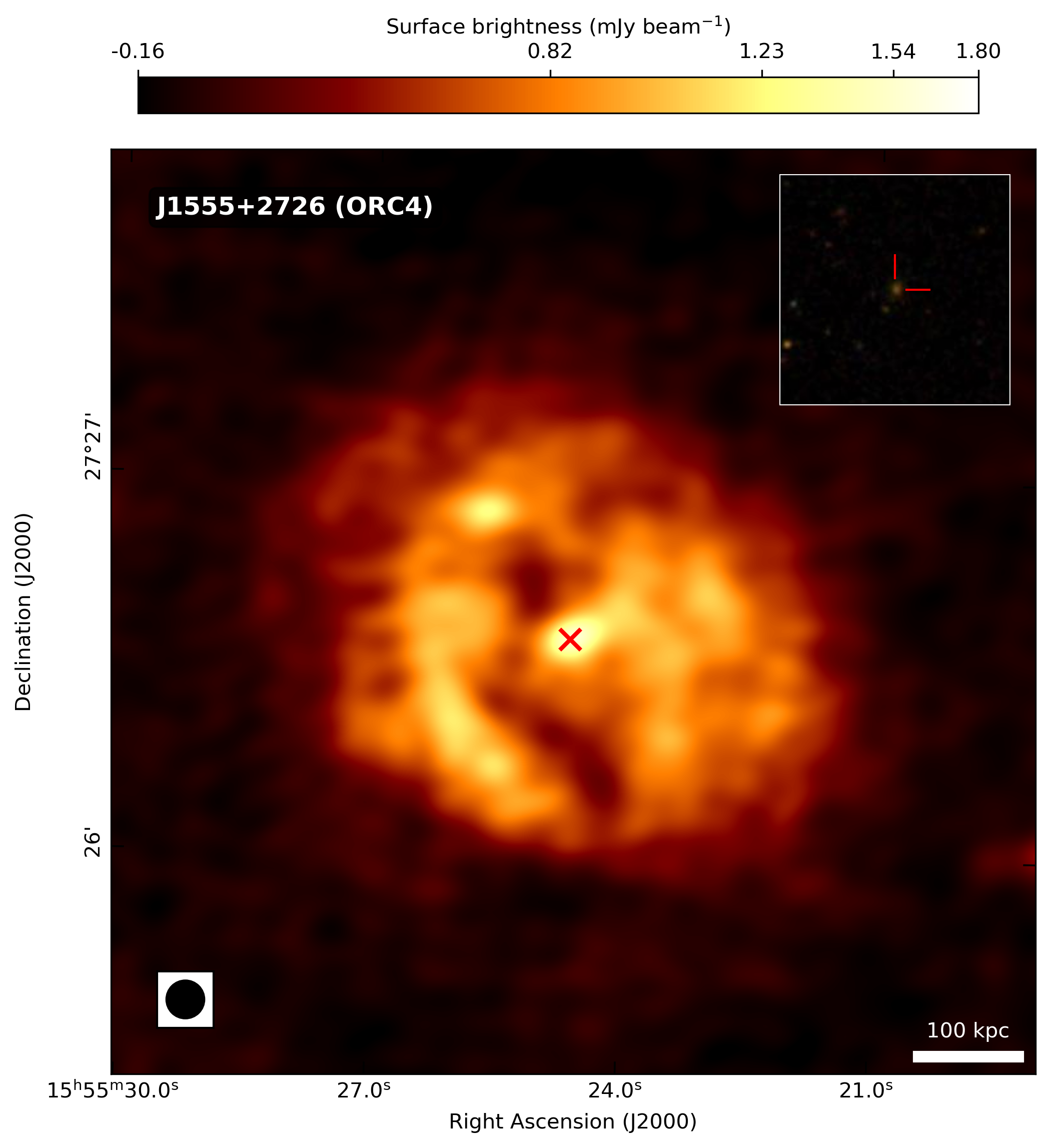} 
    \includegraphics[width=.5\textwidth]{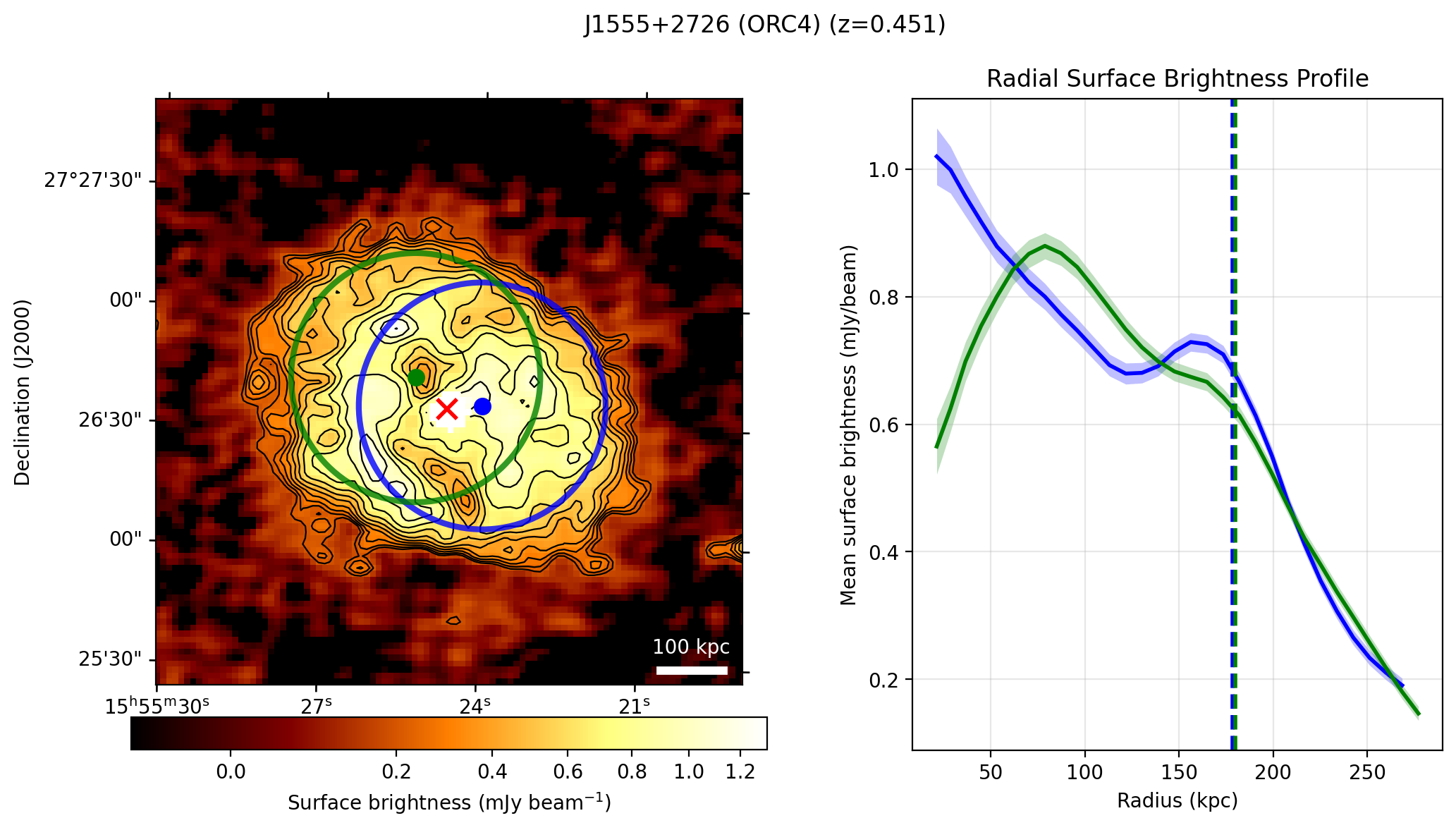}
    \caption{Same as Fig.~\ref{fig:J0823+6216}, but for DHORC J1555+2726 ("ORC 4").}
    \label{fig:J1555+2726}
\end{figure}

\paragraph{DHORC J1555+2726 "ORC 4" (Fig.~\ref{fig:J1555+2726})}
ORC 4 was first identified by~\cite{Norris2021a} and then studied in more detail in~\cite{Norris2021b}. The presence of a second ring became clear with deeper MeerKAT observations \citep{Riseley2024}. In \cite{Coil2024}, the authors found strong [O II] emission tracing ionized gas in the central galaxy of ORC 4. The observations are consistent with the infall of shock ionized gas near the galaxy following a larger, outward-moving shock that generated the radio emission. The source has an identified optical counterpart, SDSS J155524.63+272634.3, at $z=0.4512$. The source is detected in LoLSS, and the derived integrated spectral index is $\alpha = -1.3\pm0.2$.

\begin{figure}[htb!]
    \centering
    \includegraphics[width=.5\textwidth]{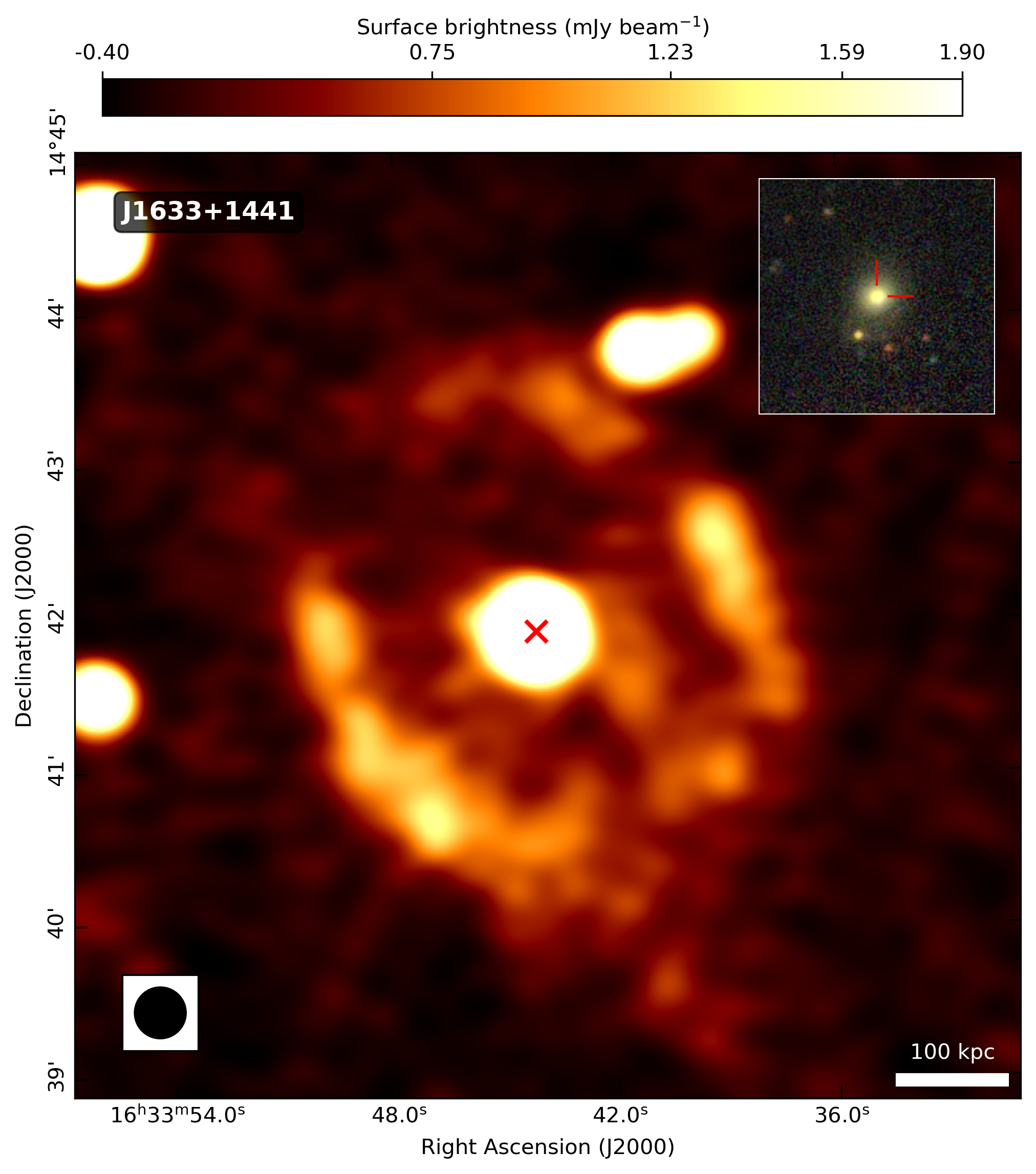}
    \includegraphics[width=.5\textwidth]{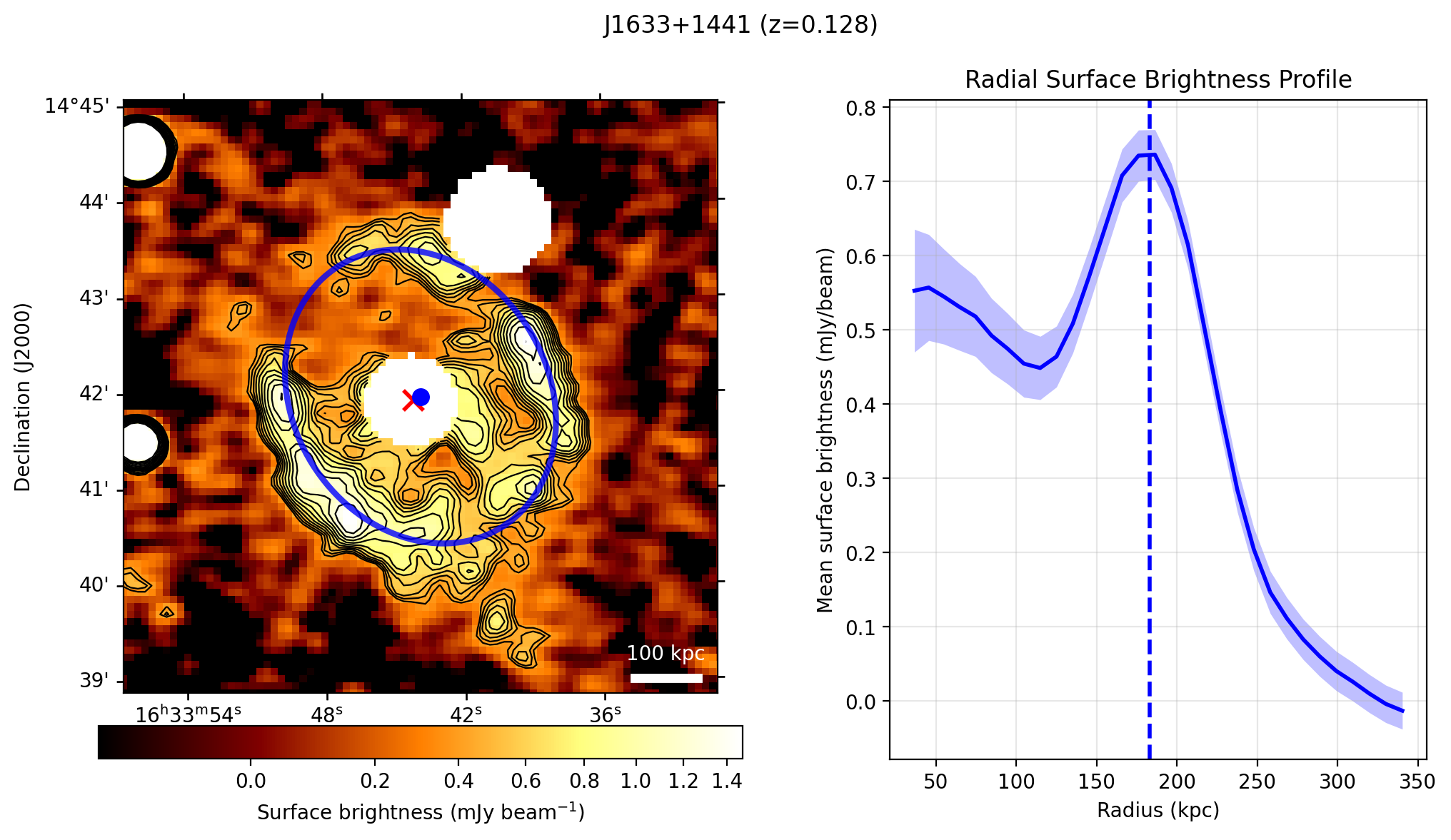}
    \caption{Same as Fig.~\ref{fig:J0823+6216}, but for ORC J1633+1441 ("Uruk-hai"). The resolution is 20\arcsec.}
    \label{fig:J1633+1441}
\end{figure}

\paragraph{ORC J1633+1441 "Uruk-hai" (Fig.~\ref{fig:J1633+1441})}
This is the best ORC of the sample in terms of morphology and clean identification of the optical counterpart. We named this source the "Uruk-hai,"\footnote{Following the name of a type of large orcs in the works of J.R.R. Tolkien.} as its apparent size is larger than that of other ORCs, with a diameter approaching 3\arcmin{} and a total source size over 4\arcmin. It has a bright central compact source that is co-located with SDSS J163344.16+144154.9, an elliptical galaxy at $z=0.1281$. The physical radius of the Uruk-hai is about 208 kpc. Unfortunately the source is not covered by LoLSS and, due to the low resolution, NVSS is too contaminated by the central source and the nearby double-lobed radio galaxy to provide a clean estimation of the spectral index.

The elliptical galaxy at the center of the Uruk-hai has been identified as a strong lens candidate in the Sloan Lens ACS (SLACS) Survey for the Masses (S4TM) lens search \citep{2017ApJ...851...48S}. This galaxy is probably distorting the light coming from a background source at $z=0.5804$. As also discussed by \citet{Bordiu2025}, any connection between an ORC of the projected size of the Uruk-hai and a galaxy-scale lens is unlikely, as the mass required to produce lensed images (in particular giant arcs) on arcmin scales is at the galaxy cluster level \citep{Meneghetti2013}. Assuming its radius as a putative Einstein radius, the mass of the lensing object would be excessive ($M_{\rm 200}\sim10^{16}$ $M_{\odot}$ assuming a standard Navarro--Frenk--White profile; \citealt{1996ApJ...462..563N}) even when considering the most massive galaxy clusters known to date, making the lensing scenario extremely unlikely.
Also, the arc-like structures of the Uruk-hai do not preserve the surface brightness across their extension, as imposed by the lensing effect, nor does the morphology seem to be compatible with a standard lens mass density distribution. It is still intriguing that two rare phenomena such as a strong gravitational lensing and an ORC coexist in the same object.

\subsection{Candidate ORCs}
Here, we list the properties of nine candidate ORCs. The images and their brightness profiles are shown in Fig.~\ref{fig: cand orc profiles}.

\paragraph{J0016+2426}
The source has a circular morphology, with extensions in three different directions. The source does not have a clear compact radio emission, and the optical counterpart closest to the ORC center is SDSS J001628.29+242610.3 ($z=0.37534$), which is a galaxy pair. The candidate ORC appears filled with radio emission, with an enhancement driven by a few localized regions at a radius of 100 kpc. The source was also detected in LoLSS and NVSS and has a power-law spectrum with $\alpha=-0.9\pm0.1$.

\paragraph{J0320+1610} The source appears very compact compared to other ORCs, but has a round shape with a linear radius of 45 kpc. There is a clear optical counterpart at the center of the circle. That is the early-type galaxy PSO J50.0394+16.1762 at $z=0.345$.  The resolution of LoTSS is insufficient to rule out the presence of a central point radio source. FIRST does not cover the target and VLASS does not show any point source corresponding to the target. LoLSS data are not available for this target. The spectral index obtained from LoTSS and NVSS (144 and 1400 MHz) is $\alpha=-0.72\pm0.01$, similar to what found at lower frequency in GLEAM-X \citep[$\alpha = -0.84$;][]{Ross2024}.

\paragraph{J0324+8322}
The source has an elliptical shape but most of the flux density comes from the central/north region. The edges are sharp but the brightness enhancement is dominated by a localized region in the north. There is no apparent point source within the candidate ORC boundaries, and we associated as optical counterpart the galaxy closest to the ellipse's center. The associated galaxy is PSO J051.2103+83.3666 ($z=0.3655$), whose distance gives a mean source radius of 174 kpc. However, there are other possible optical counterparts in the vicinity such as PSO J051.1879+83.3702, whose redshift is unknown, but likely higher. The source was clearly detected in LoLSS and NVSS, yielding a power-law spectrum with spectral index of $\alpha=-0.97\pm0.05$.

\paragraph{J0801+5544}
The source has a round shape with hints of brightness excess on the edges. However, most of its emission comes from the central region of the source. A number of substructures are present in the source that might be a superposition of multiple, circularly shaped structures. We cannot identify a compact radio source within the candidate ORC boundaries, and we tentatively associated the optical counterpart SDSS J080103.48+554424.8 ($z=0.8$). Assuming the correct redshift estimation, the source is more luminous than other ORCs or candidate ORCs, with a luminosity of $L_{144} = (4.4 \pm 0.1) 10^{26}$~W~Hz$^{-1}$. The source has a radius of 210 kpc. The source was also detected in LoLSS and NVSS, which yields an integrated spectral index of $\alpha=-1.4\pm0.1$.

\paragraph{J0940+6028}
This source has an elliptical shape, and the edge emission is only partly enhanced. The mean radius of the source is 160 kpc. The region enclosed by the ellipses is filled with radio emission that seems structured, but higher resolution images are needed to assess its morphology. A bright spot of radio emission is visible north of the center of the ellipse. We used this location to tentatively identify an optical counterpart, that is, SDSS J094051.23+602846.7 ($z=0.53327$). No compact source was detected in FIRST, nor in VLASS. The radio spectrum of this source is consistent with a power law between 54 and 1400 MHz and a slope of $\alpha=-1.18\pm0.03$.

\paragraph{J1436+4832}
This source is colocated with a known galaxy group, and it may be similar to the Cloverleaf ORC \citep{Bulbul2024}. We could identify a point radio source close to the source center that is associated with a bright elliptical galaxy (SDSS J143619.42+483210.5; $z=0.19115$). The radio emission shows a circular brightness enhancement at the radius of 104 kpc; the circle is centered on the bright point source. The source is clearly detected in LoLSS and NVSS. The derived integrated spectrum is a power law with a spectral index of $\alpha=-1.14\pm0.03$. The emission from the compact source is subdominant and can be neglected when measuring the flux densities for the integrated spectrum.

\paragraph{J1458+4534}
This source is characterized by a partial thin and disrupted ring. We were not able to not identify any optical counterpart close to the geometrical center of the ring, and no radio point sources were detected in the region. For this reason we classify this source as a "candidate" ORC. We derived a low-frequency spectral index using LoLSS and LoTSS, obtaining $\alpha = -1.73 \pm 0.06$.

\paragraph{J1559+2734}
This candidate ORC shows a half-circle of enhanced emission with $r = 0.24\arcmin$. The object was not detected in LoLSS, but it was detected in NVSS close to the $3\sigma$ cut. The derived spectral index is $\alpha = -0.77\pm0.07$. There is no clear central compact radio source nor any known optical counterpart in the vicinity of the circle's center.

\paragraph{J1608+6123}
This candidate ORC is elongated, with an average radius equal to 93 kpc. The source appears to be filled with radio emission that is edge-brightened. There is no clear evidence of a point-like radio source near the center, and no detection is reported in FIRST and VLASS. We tentatively associate the emission with the closest visible galaxy (SDSS J160833.95+612356.8; $z=0.34$). The source is detected close to the $3\sigma$ cut in NVSS, and the extracted spectral index is a power law with $\alpha=-0.92\pm0.08$.

\begin{figure}[!t]
    \centering
    \includegraphics[width=\linewidth]{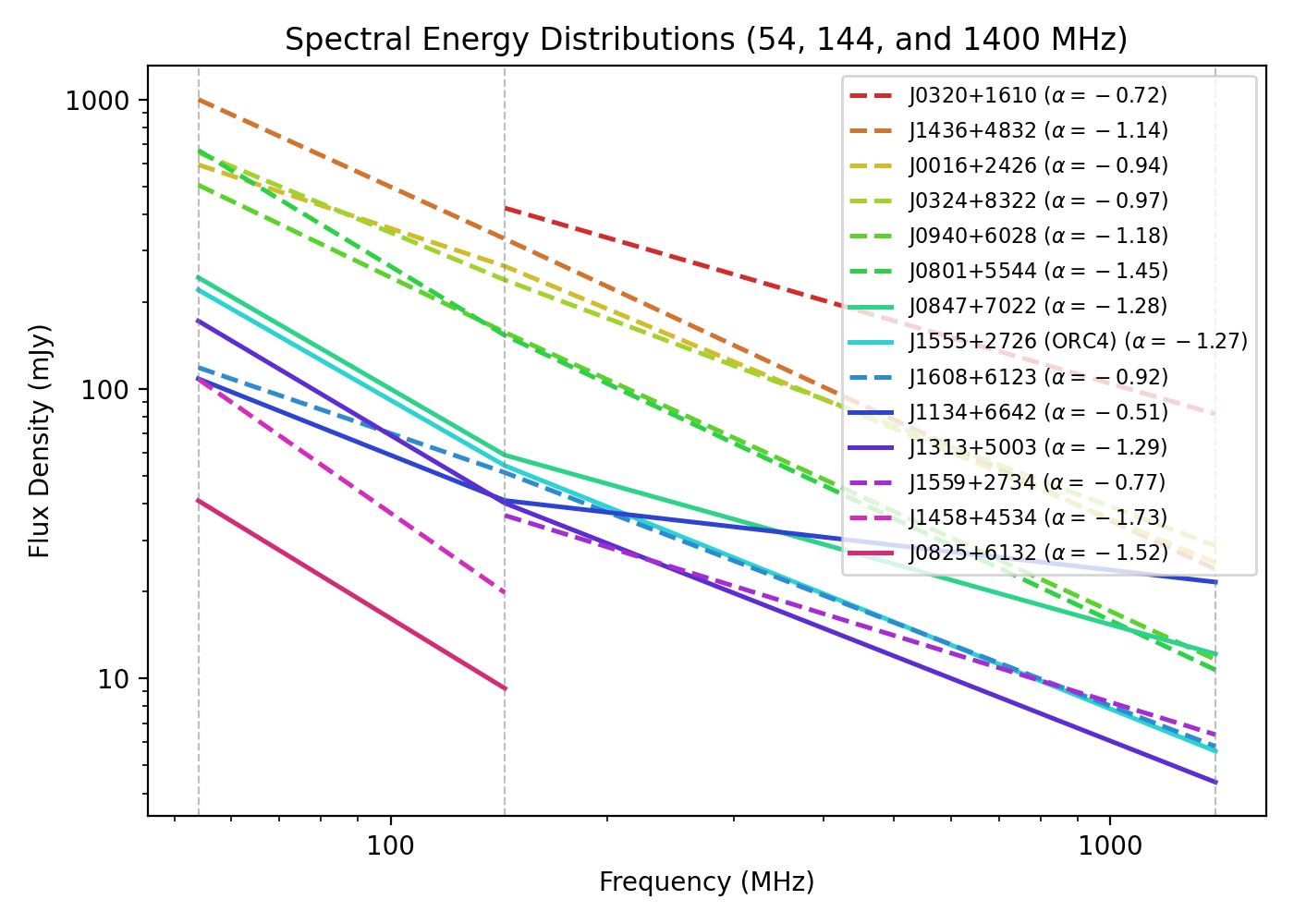}
    \caption{Radio spectra for ORCs where at least one other frequency than LOFAR 144 MHz was available. Dashed lines are for candidate ORCs. The three vertical lines correspond to the three surveys: LoLSS (54 MHz), LoTSS (144 MHz), and NVSS (1400 MHz).}
    \label{fig:sed}
\end{figure}

\section{Discussion}
\label{sec:discussion}

Given the growing number of sources that fall into the ORC class, we can start deriving some global properties in order to try and understand their origin. 

\subsection{Host galaxies}
Our data confirm that all bona fide ORCs have a large elliptical galaxy as an optical counterpart. This counterpart can be either radio-detected or not. However, some of our candidate ORCs do not show the presence of a central elliptical galaxy, because (i) of a large displacement from the ORCs' centres, (ii) the source is not visible in the optical surveys, or (iii) of a misclassification of the radio emission. We also confirm that the optical counterparts of our close ($z<0.5$) ORCs and DHORCs show either signs of interaction or lie in an overdense environment. This is clear for ORC~0823+6216 and ORC~1633+1441, which both have a number of neighboring galaxies; for DHORC~0847+7022, whose host galaxy lies in a small group; and for ORC~1134+6642 and ORC~1222+6436, whose hosts appear to have a perturbed morphology. The redshift of the other ORCs is too high to draw any conclusions regarding their dynamical states.

\subsection{Comparison with known ORCs}
The spectral indices of the known ORCs lie mostly between -1 and -1.7 \citep{Norris2021a, Koribalski2025, Hota2025}. Also, the ORCs and candidate ORCs in the sample presented in this work have a steep spectrum. Their spectral energy distribution is shown in Fig.~\ref{fig:sed}. Such a steep spectrum implies that the CRs in the ORCs are not actively being powered, but they are only radiating and thus losing energy. Hence, any local acceleration mechanism must be inefficient or must have already ceased. To investigate further, we collected the spectral indices of ORCs and candidate ORCs from this work and from the literature. When possible, we tried to isolate the spectral index of the shell, removing the contribution of the radio-loud host galaxy. We caution that the spectral indices taken from the literature are extracted at nonuniform frequencies.

\begin{figure}[!t]
    \centering
    \includegraphics[width=\linewidth]{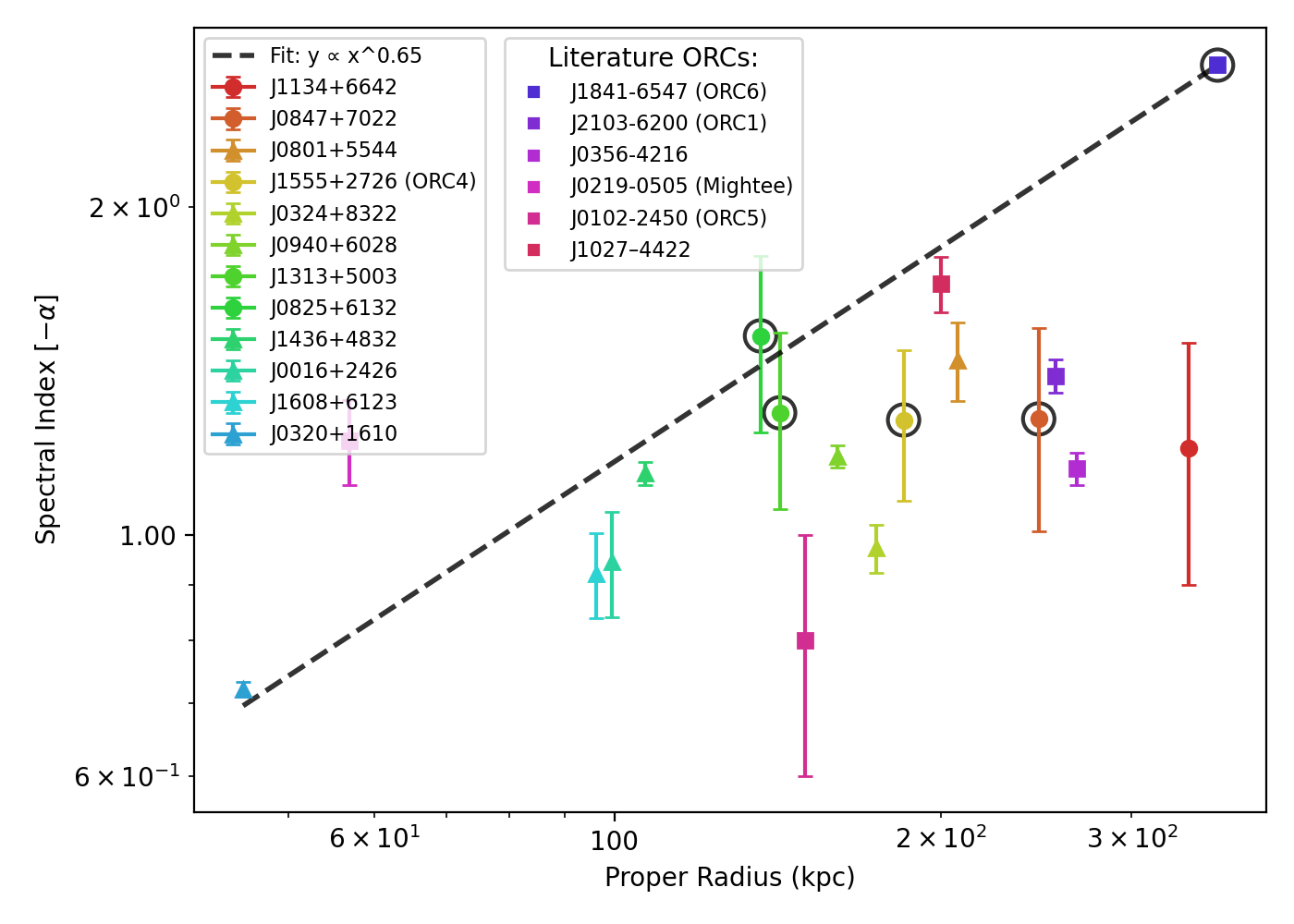}
    \caption{Integrated spectral index of ORCs (circles) and candidate ORCs (triangles) in our sample plotted against the physical radius of the source. Markers within a circle are for DHORCs. The dashed line shows a log-log linear regression. We included the following sources from the literature: J1841-6547 (ORC6: \citealt{Koribalski2025}), J2103-6200 (ORC1; \citealt{Norris2021a}), J0356-4216 \citep{Taziaux2025}, 0219-0505 (Mightee; \citealt{Norris2025}), J0102-2450 (ORC5; \citealt{Koribalski2021}), and J1027–4422 \citep{Koribalski2024a}.}
    \label{fig:radiusspidx}
\end{figure}

Next, we plotted their spectral indexes against their physical sizes in Fig.~\ref{fig:radiusspidx}.
With the notable exception of ORC J0219-0505 (MIGHTEE), we find that all known ORCs (as well as our candidates) lie in the same region of the size--spectral index plot. The plot underlines the absence of ORCs that are both small in size and have a steep spectra. This is in agreement with the idea of an expanding shock that either loses energy after expansion or transition to a region with fewer CRes to re-energize. Even a mild shock should have an integrated spectral index of $\alpha \sim -1$, so at least in some cases the steep spectrum cannot be simply due to the presence of aged electrons. However, literature ORCs with very steep spectra were identified based on data at higher frequencies, therefore biasing the comparison if the spectrum is curved. Other sources with steep spectra are double-headed ORCs, which may point toward a different nature for these sources. For instance, they could be dying lobes of radio galaxies with peculiar morphologies.

\subsection{Comparisons with simulations}

Several formation pathways for ORCs have been proposed that can qualitatively explain their morphology. In particular, three mechanisms have been explored with simulations so far: (i) merger shocks in massive galaxies and galaxy groups \citep{Dolag2023a,Ivleva2026}; (ii) AGN-injected cosmic rays \citep[e.g.][]{Nolting2023, Shabala2024}; and (iii) starburst winds \citep{Coil2024}.

So far, only the merger-driven shock scenario has been tested in magnetohydrodynamic simulations with self-consistent modeling of cosmic-ray populations and magnetic fields \citep{Ivleva2026}. There, the authors conducted a cosmological zoomed-in simulation of an assembling galaxy group, which undergoes a major merger event in its late evolution. This triggers several shock waves in the CGM, which efficiently accelerate cosmic-ray electrons and hence become visible in the radio through synchrotron radiation. While this model is able to reproduce the morphology and spectral index of ORCs, the total emitted power at \mbox{150 MHz} is at least three orders of magnitude lower than in observed ORCs. At the same time, the polarization intensity is systematically overpredicted by about 10\%, though this is a relatively small discrepancy and could be well within observational uncertainties. According to \cite{Ivleva2026}, these inconsistencies could be caused by a combination of missing cosmic-ray injection mechanisms and magnetic-field amplification processes. In particular, fossil cosmic-ray populations injected by AGN and stellar feedback events should contribute to the budget of nonthermal electrons \citep[see][]{Shabala2024} aside from diffusive shock acceleration, which was the only injection mechanism in their model. In addition, the evolution of the magnetic-field strength and subsequent radio emission can be significantly boosted by plasma processes at the shock, which are not yet implemented in cosmological simulations.

\begin{figure*}
    \centering
    \includegraphics[width=.33\linewidth]{1222+6436_low.png}
    \includegraphics[width=.33\linewidth]{1150+2449_high.png}
    \includegraphics[width=.33\linewidth]{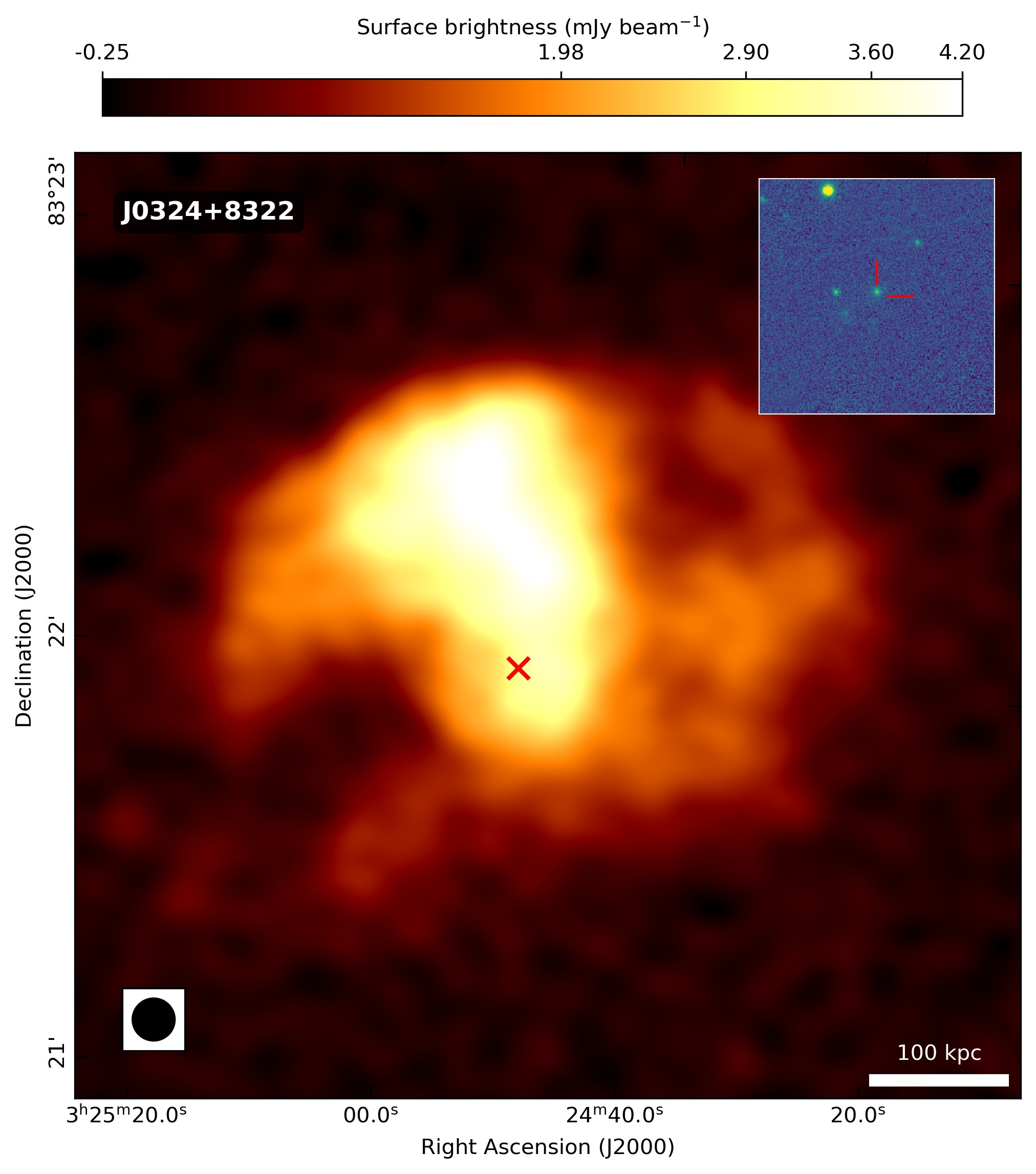}\\
    \includegraphics[width=.33\linewidth]{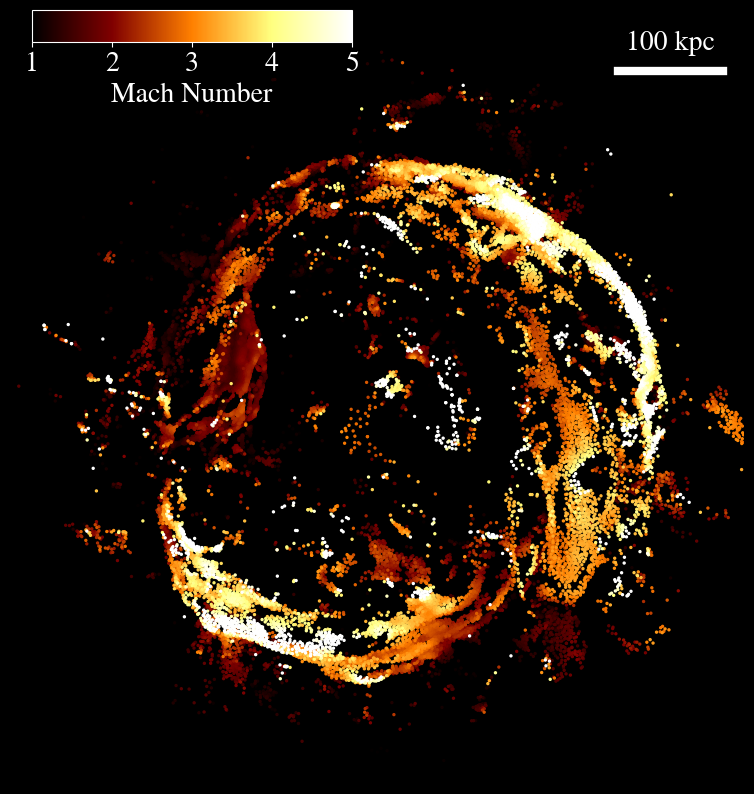}
    \includegraphics[width=.33\linewidth]{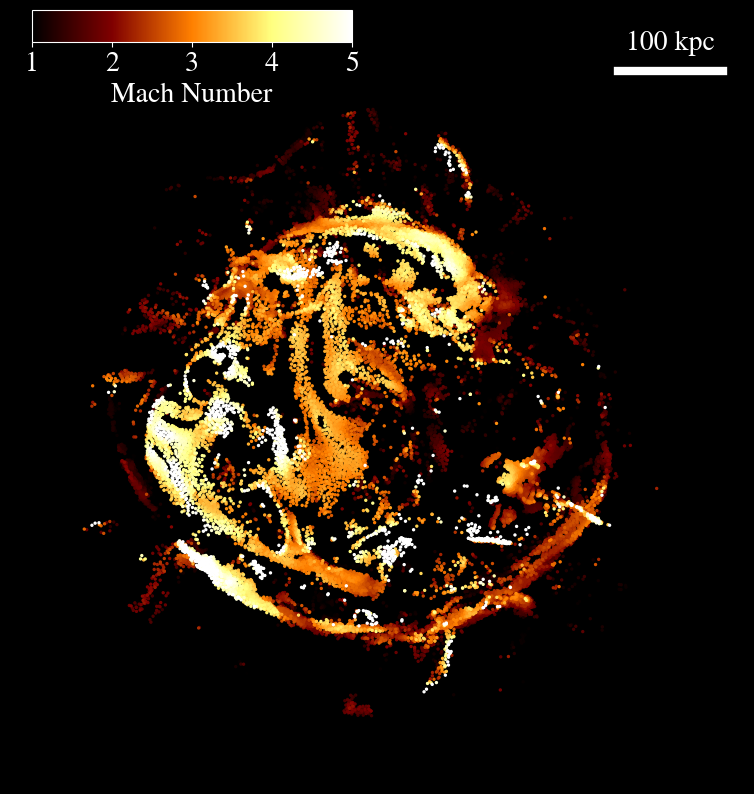}
    \includegraphics[width=.33\linewidth]{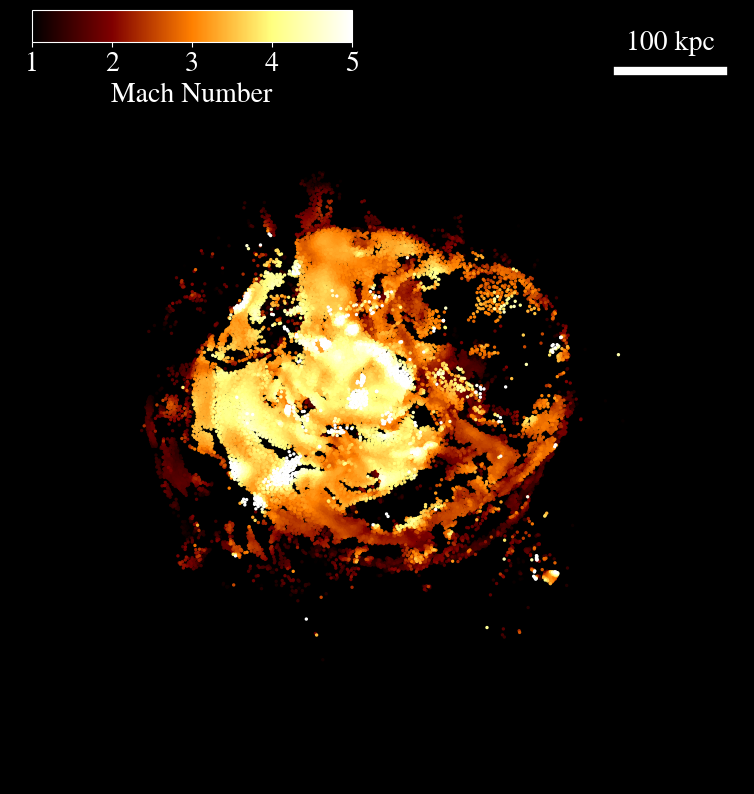}
    \caption{Morphological comparison between a set of targets presented in this paper (top row) with a simulation by \citet[bottom row]{Ivleva2026}. The red cross in the observed objects marks the position of the optical counterpart shown in the inset at the top right corner of each panel, while the beam size is indicated at the bottom left. The color in the images of the simulation displays the local Mach number of a shock, which is triggered by a major merger event inside the galaxy group. The morphology and size of the observed objects fits the simulated counterpart well, which in each of the three cases is shown at a different projection and time.}
    \label{fig: simul}
\end{figure*}

In Fig.~\ref{fig: simul}, we compare a set of our sources shown in the top row to the simulated ORC published by \cite{Ivleva2026} in the bottom row. In the simulation images, the color bar shows the local sonic Mach number of the shock structure, while the scale in the top right indicates a physical size of \mbox{100 kpc}. Each column in the bottom row shows the same simulated object, but viewed at different projections and time steps. The merger scenario inherently contains a preferred axis, namely the merger axis along which the two galaxies coalesce inside the group environment. When the line of sight is perpendicular to that axis, the emerging shock waves appear to form a circle around the merger site. The bottom left panel in Fig.~\ref{fig: simul} shows such a line of sight in the simulation. Evident when comparing to the panel above, which displays an example from our observational sample with the clear ring structure inherent to ORCs, this scenario can easily reproduce the morphology and extent of such objects. However, since this particular formation scenario is not isotropic, this has interesting consequences for the diversity of observational counterparts. The middle and right panels in the bottom of Fig.~\ref{fig: simul} demonstrate two such examples, where the observer positions are varied such that the line of sight is tilted compared to the fiducial ORC projection in the leftmost panel. As found in our sample, more complex morphologies are also possible, where the radio ring appears to be partially filled toward the center (middle column) or exhibits entirely centrally dominated emission (right column).

Since the shock waves quickly expand radially into the outskirts of the CGM, this is also naturally reflected in the physical size of the ORC-like structure. As noted above, the bottom row of Fig.~\ref{fig: simul} displays the simulated counterpart at different times. Specifically, the middle and right panels show the object about 10 and 20 Myr earlier, respectively, compared to the leftmost image. Thus, the size of ORCs can quickly vary by orders of hundreds of kiloparsecs in just a few dozen megayears, effectively explaining the rarity of such objects, since they only maintain their state for a relatively short time. In summary, the variety in the appearance and size of one single simulated object displayed in the bottom row of Fig.~\ref{fig: simul} ---as well as the matching morphology with various targets presented in this paper--- suggests that the now emerging diversity of observed peculiarities in diffuse radio emission does not necessarily require a variety of formation scenarios. Instead, they could be realizations of the same handful of mechanisms, where the morphological variance is induced by the line of sight and time of the observation.

\section{Conclusions}
\label{sec:conclusions}

We significantly advanced our knowledge of ORCs by expanding the sample of known sources using LoTSS data. In particular, we identified five ORCs, nine candidate ORCs, and two newly discovered double-headed ORCs, adding to the diversity of the ORC population. Our search combined catalog selection and visual inspection, supplemented by a multifrequency spectral analysis, to explore common characteristics such as morphology, spectral index, and associations with optical counterparts.

Our results confirm that ORCs are a heterogeneous population, often associated with massive elliptical galaxies and exhibiting varied morphologies including single and double ring-like structures with sharp edges and steep radio spectra ($\alpha$ between $-0.7$ and $-1.7$).
There is no clear correlation between the physical size of the source and the spectral index, but steep spectrum ORCs are preferentially those with the largest physical size, as one would expect if the resupply of CRe has stopped or the DSA injects CRe at increasingly weak shocks. At least in a test-particle DSA, the integrated synchrotron spectra steepen with decreasing Mach number.

All bona fide ORCs have a large elliptical galaxy as an optical counterpart, but for some of our candidate ORCs we were not able to identify a central elliptical galaxy, because of a large displacement from the ORCs' centers, because the host is not visible in the optical surveys, or because of a misclassification of the radio emission. Interaction signatures in the host galaxies point toward dynamic environments influencing ORC formation. A comparison with current theoretical models suggests that no single proposed mechanism can fully explain the observed properties and diversity of ORCs.


\begin{acknowledgements}
FDG, TP and GDG acknowledge support from the ERC Consolidator Grant ULU 101086378. MB acknowledges funding by the Deutsche Forschungsgemeinschaft (DFG) under Germany's Excellence Strategy -- EXC 2121 ``Quantum Universe" --  390833306 and the DFG Research Group "Relativistic Jets" FOR5195 – project number 443220636. AI acknowledges support by the COMPLEX project from the European Research Council (ERC) under the European Union’s Horizon 2020 research and innovation program grant agreement ERC-2019-AdG 88267.
CS acknowledges the support by the Italian Ministry of University and Research (grant FIS2023-01611, CUP C53C25000300001).
LMB is supported by NASA through grant 80NSSC24K0173 and NSF through grant AST-2510951. MJH thanks the UK STFC for support [ST/Y001249/1].

We thank Massimo Meneghetti for the useful discussions on this work.

LOFAR is the Low Frequency Array designed and constructed by ASTRON. It has observing, data processing, and data storage facilities in several countries, which are owned by various parties (each with their own funding sources), and which are collectively operated by the LOFAR ERIC under a joint scientific policy. The LOFAR resources have benefited from the following recent major funding sources: CNRS-INSU, Observatoire de Paris and Université d'Orléans, France; BMBF, MIWF-NRW, MPG, Germany; Science Foundation Ireland (SFI), Department of Business, Enterprise and Innovation (DBEI), Ireland; NWO, The Netherlands; The Science and Technology Facilities Council, UK; Ministry of Science and Higher Education, Poland; The Istituto Nazionale di Astrofisica (INAF), Italy.

This research made use of the Dutch national e-infrastructure with support of the SURF Cooperative (e-infra 180169) and the LOFAR e-infra group. The Jülich LOFAR Long Term Archive and the German LOFAR network are both coordinated and operated by the Jülich Supercomputing Centre (JSC), and computing resources on the supercomputer JUWELS at JSC were provided by the Gauss Centre for Supercomputing e.V. (grant CHTB00) through the John von Neumann Institute for Computing (NIC).

This research made use of the University of Hertfordshire high-performance computing facility and the LOFAR-UK computing facility located at the University of Hertfordshire and supported by STFC [ST/P000096/1], and of the Italian LOFAR IT computing infrastructure supported and operated by INAF, and by the Physics Department of Turin university (under an agreement with Consorzio Interuniversitario per la Fisica Spaziale) at the C3S Supercomputing Centre, Italy.

This research is part of the project LOFAR Data Valorization (LDV) [project numbers 2020.031, 2022.033, and 2024.047] of the research programme Computing Time on National Computer Facilities using SPIDER that is (co-)funded by the Dutch Research Council (NWO), hosted by SURF through the call for proposals of Computing Time on National Computer Facilities.
\end{acknowledgements}

\bibliographystyle{aa}
\bibliography{library}

\onecolumn
\begin{appendix}
\section{Images and brightness profiles of candidate ORCs}

\begin{figure*}[!htb]
\centering
\includegraphics[width=.43\textwidth]{
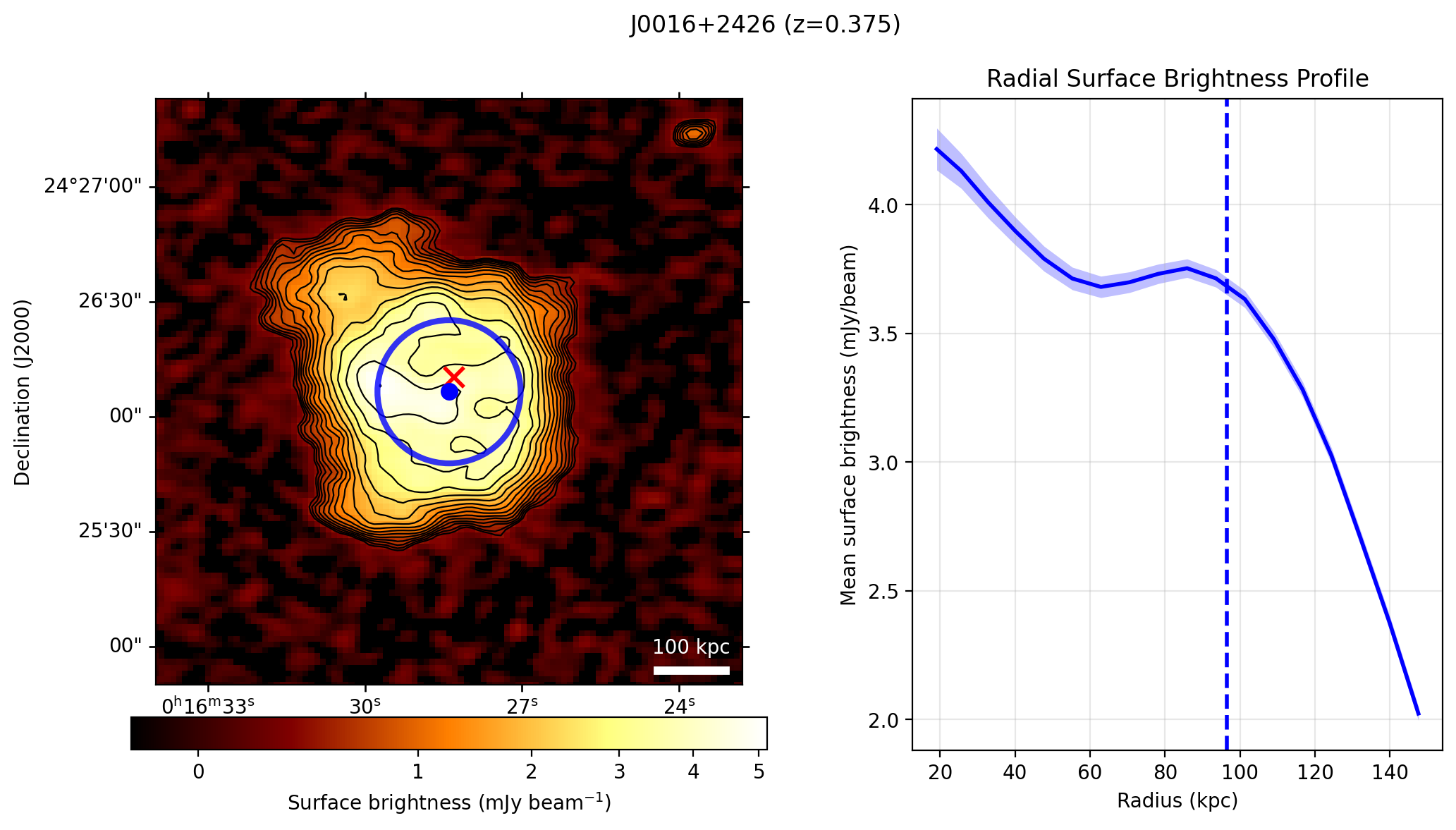}
\includegraphics[width=.43\textwidth]{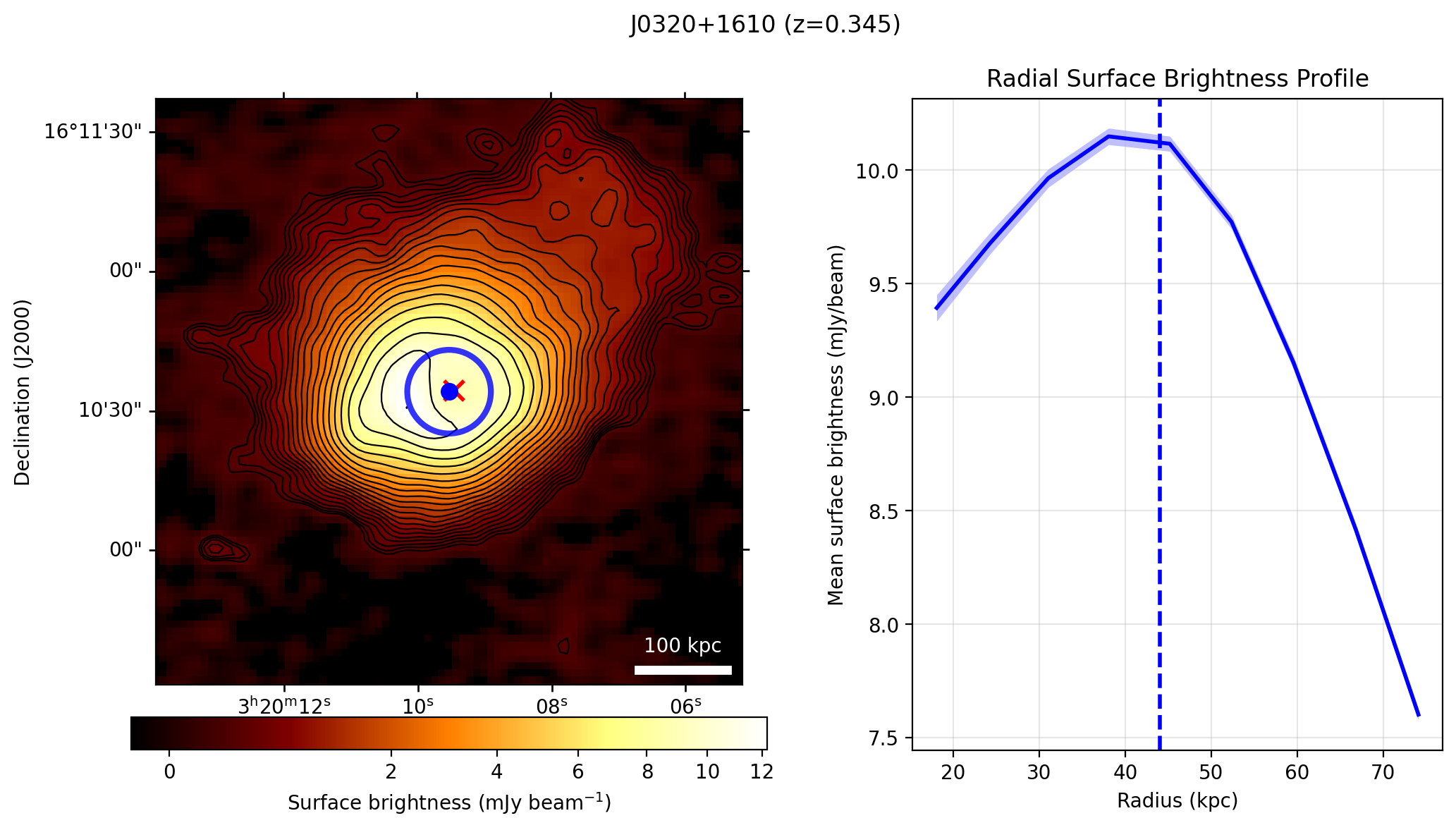}
\includegraphics[width=.43\textwidth]{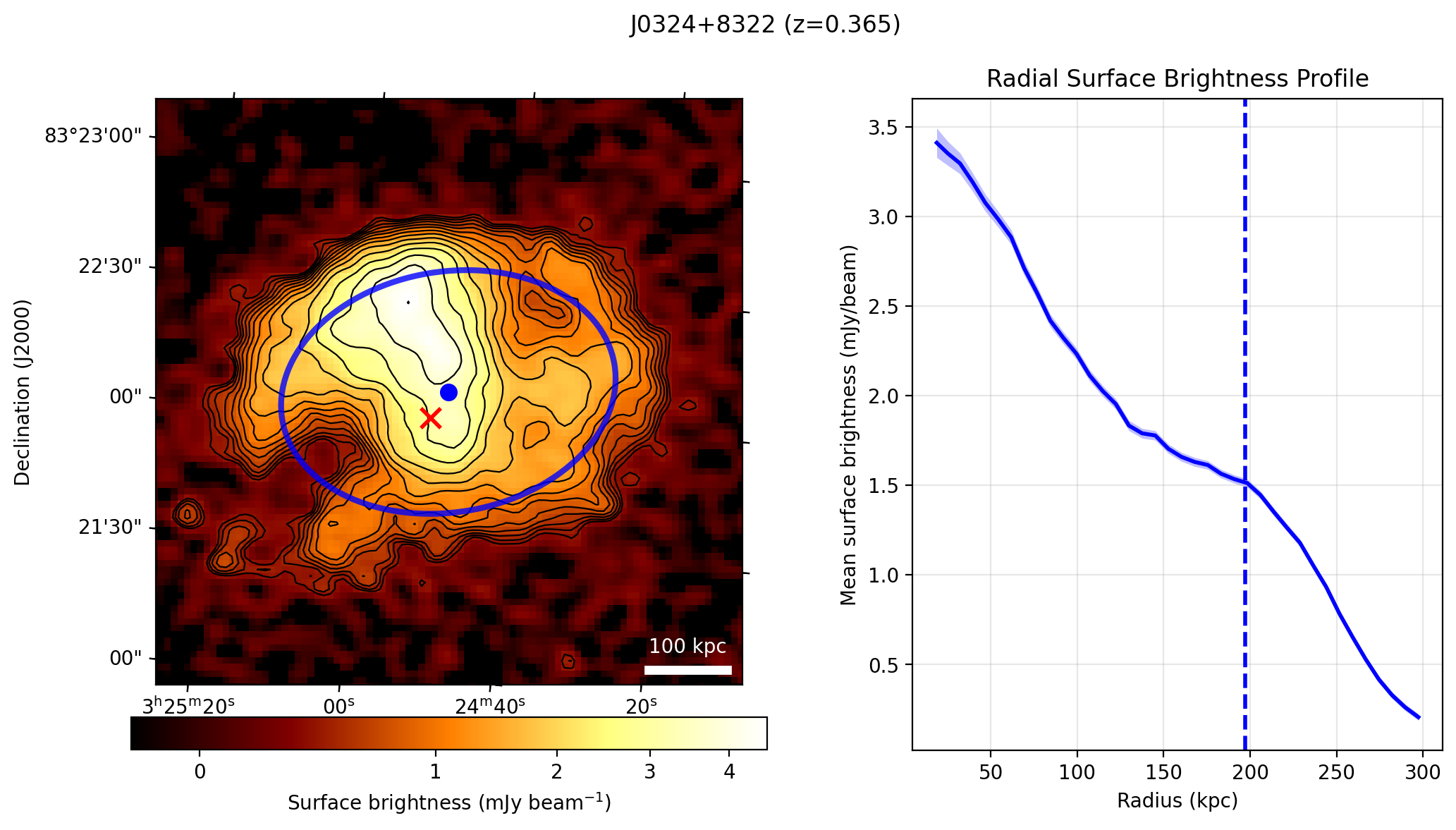}
\includegraphics[width=.43\textwidth]{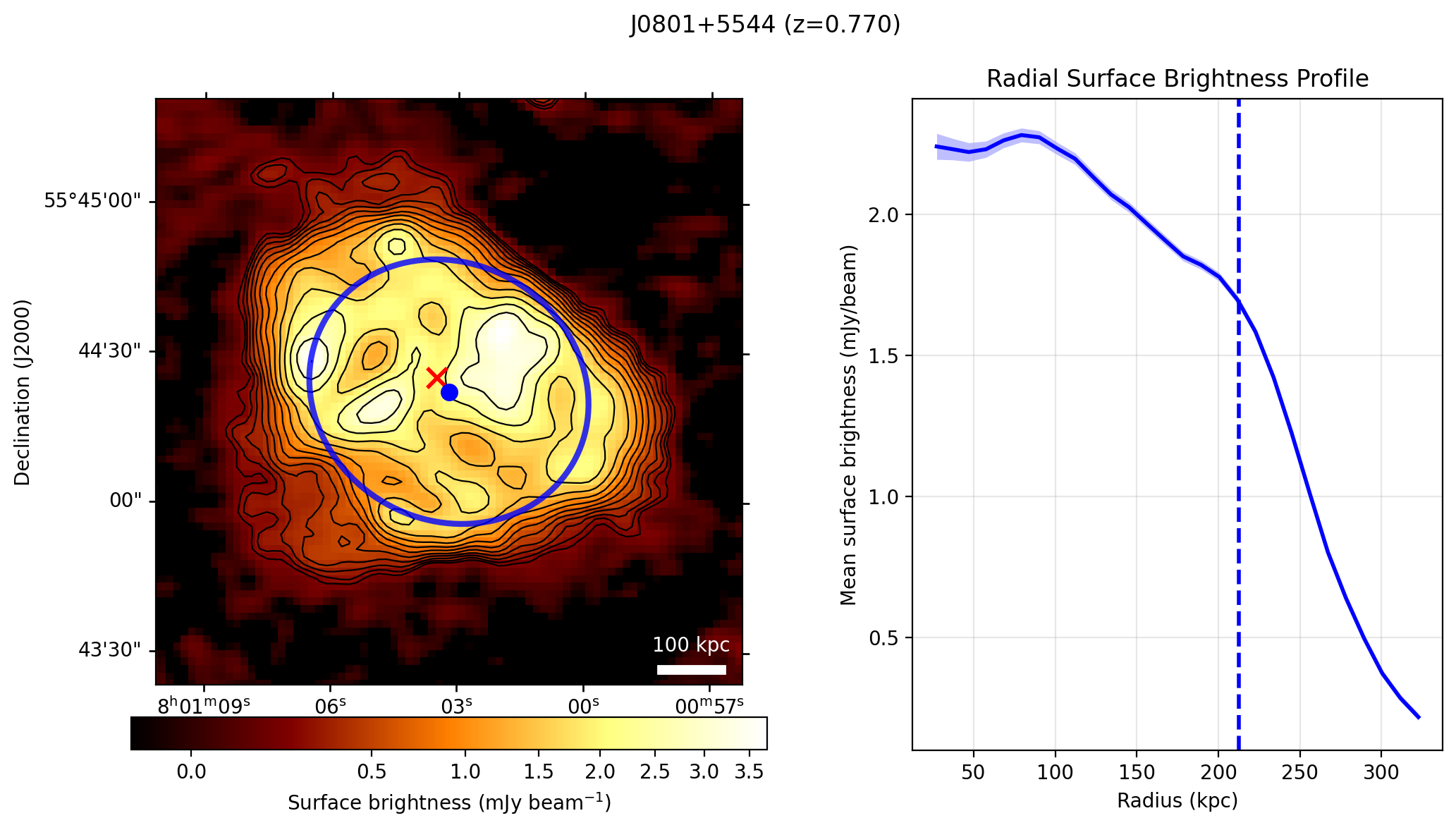}
\includegraphics[width=.43\textwidth]{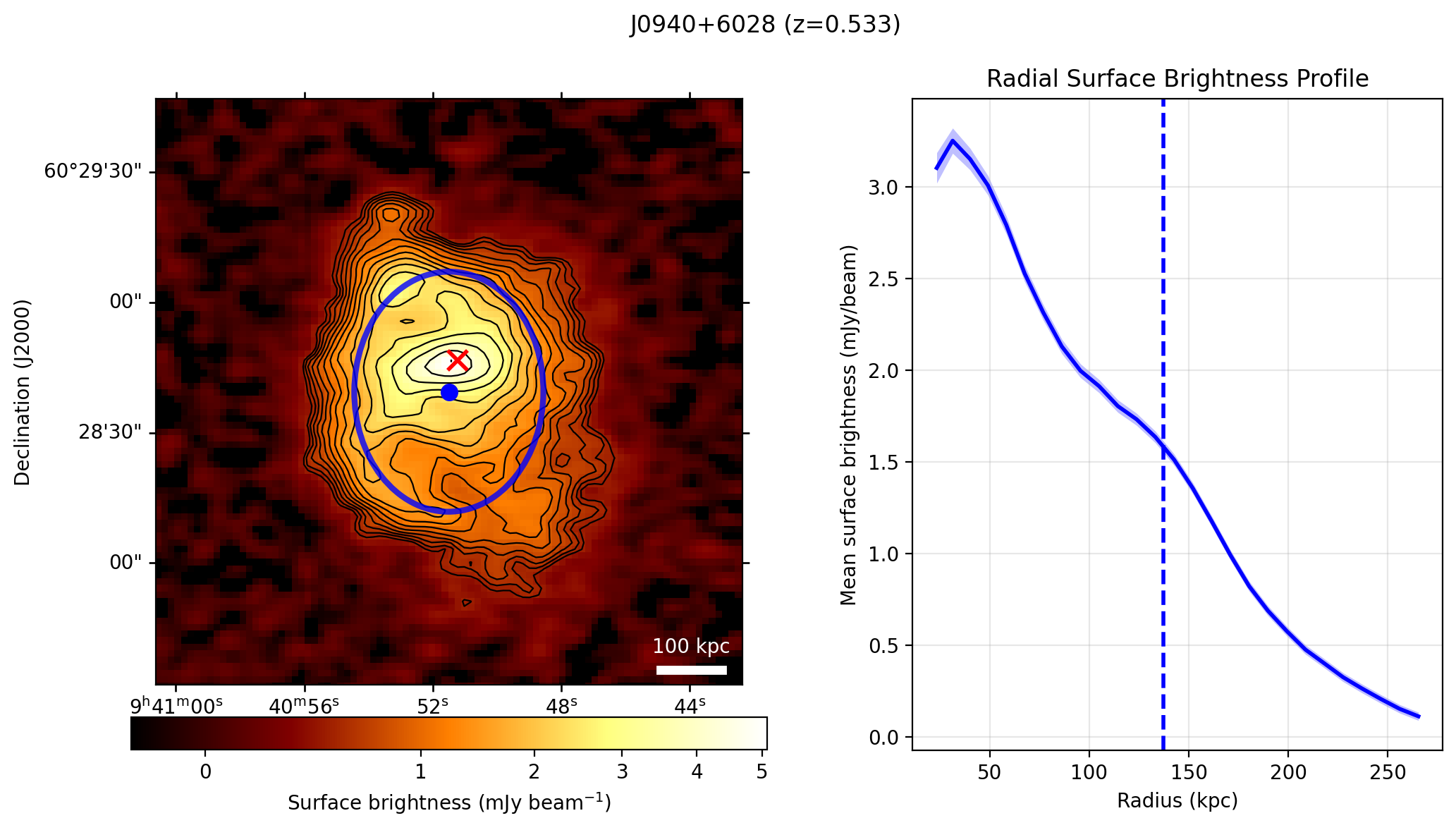}
\includegraphics[width=.43\textwidth]{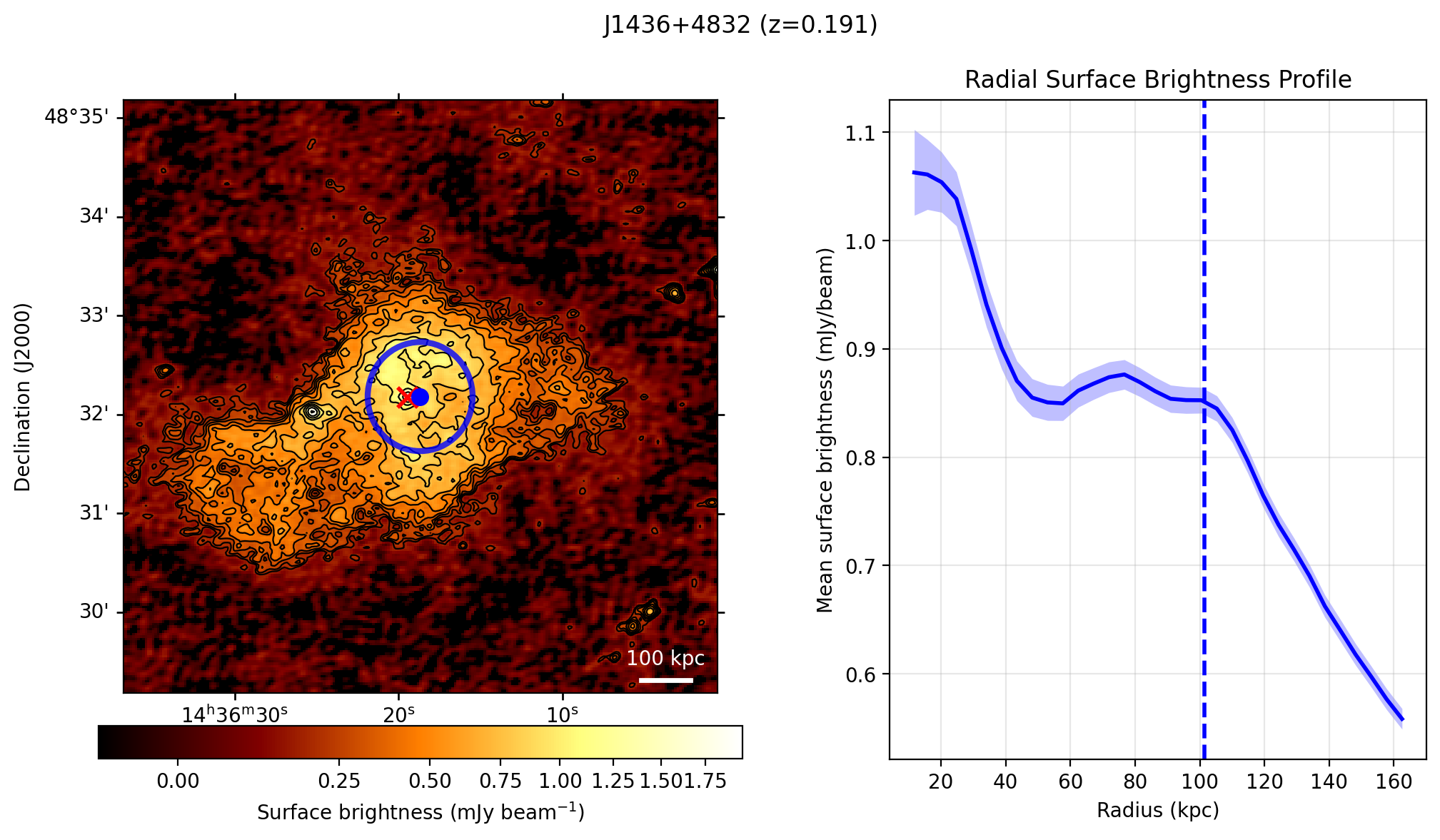}
\includegraphics[width=.43\textwidth]
{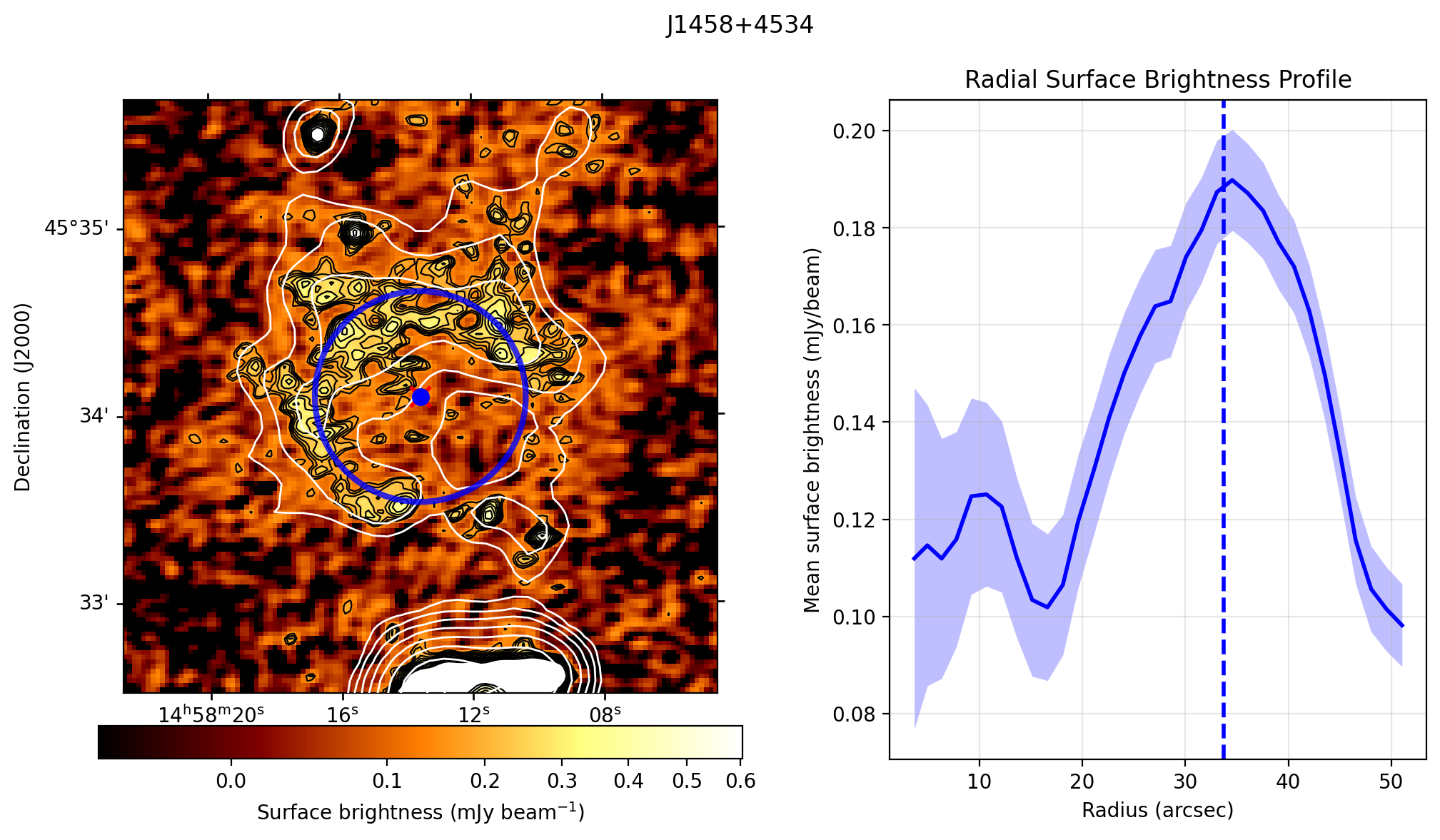}
\includegraphics[width=.43\textwidth]
{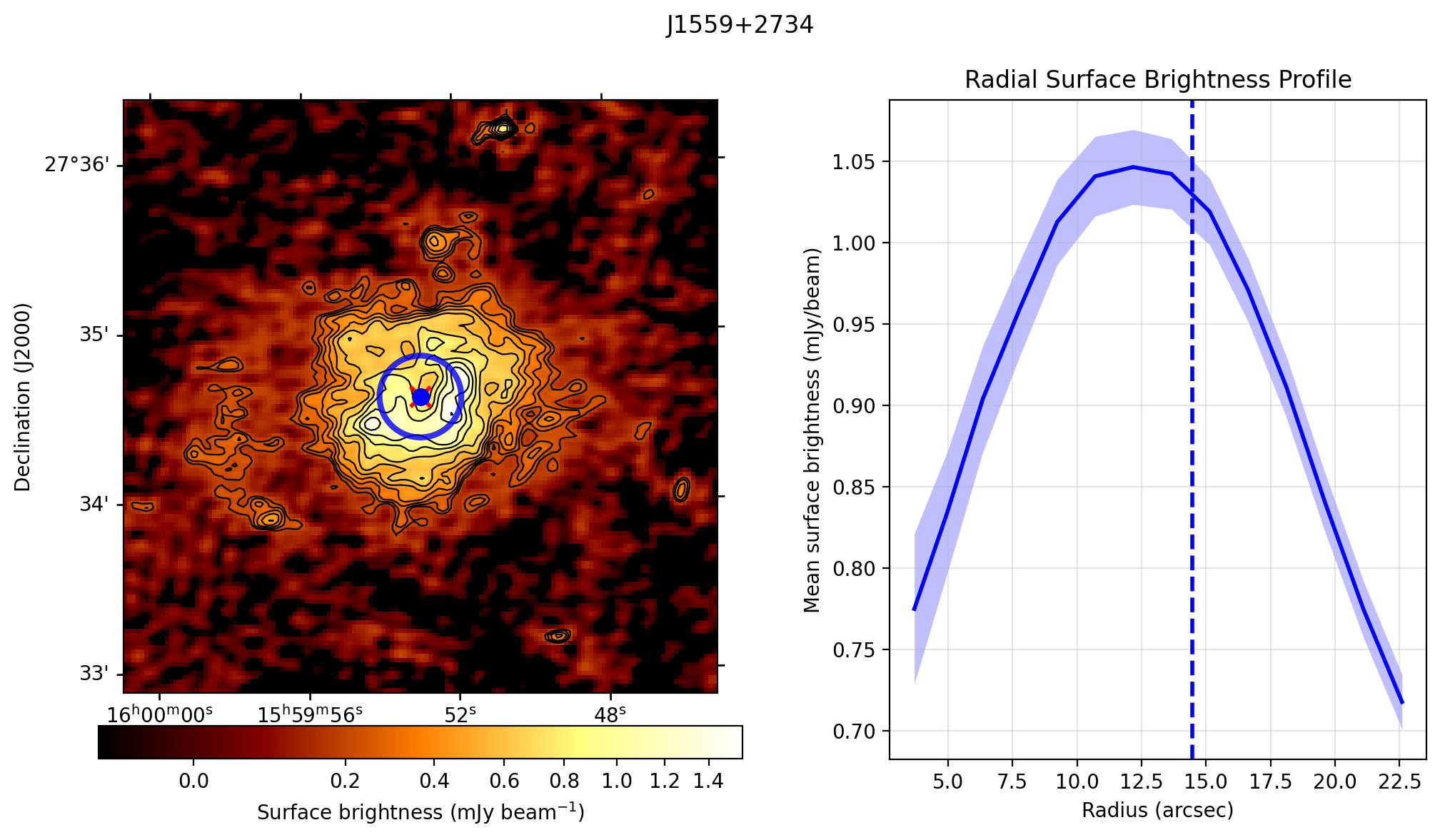}
\includegraphics[width=.43\textwidth]{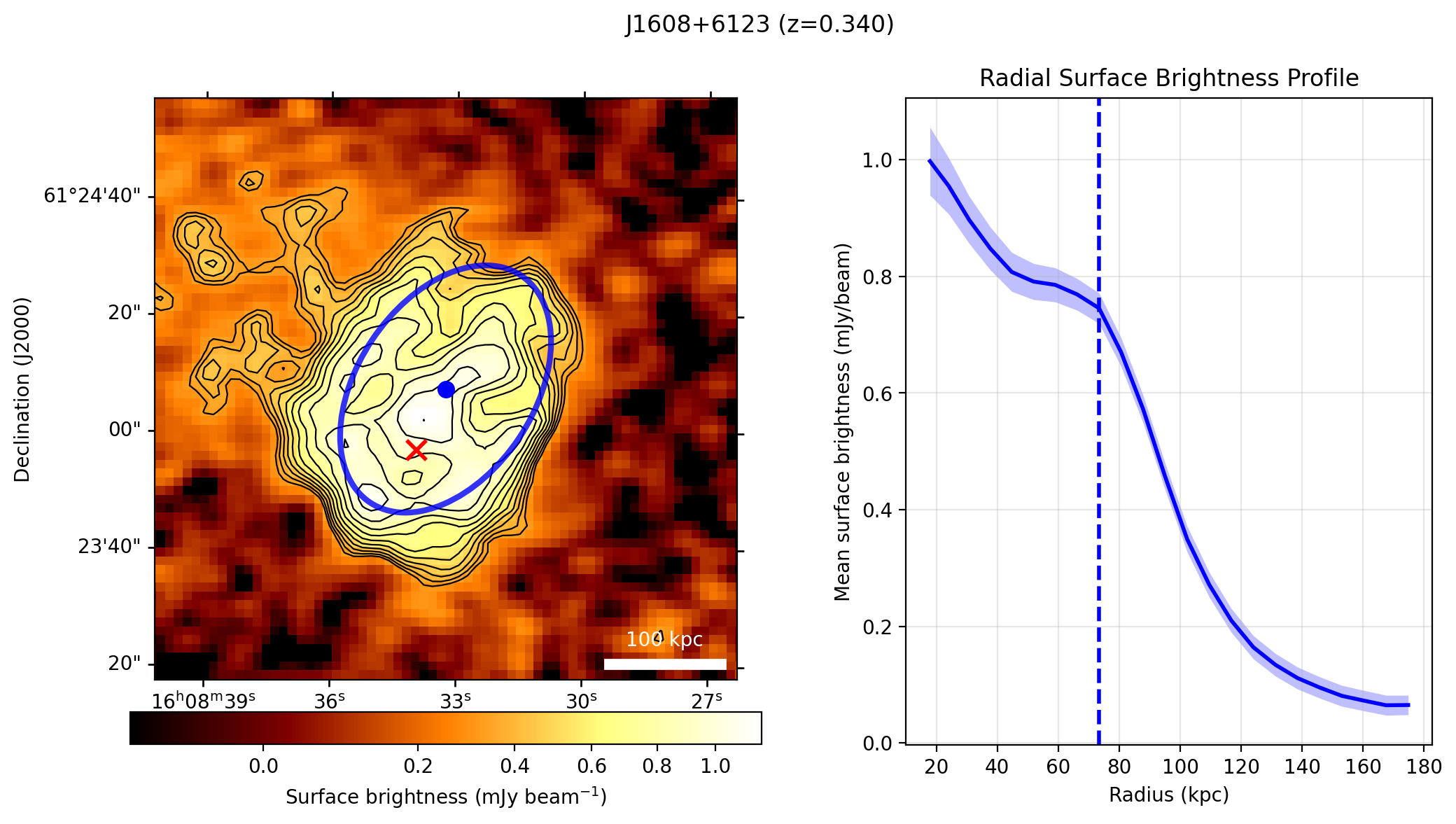}
    \caption{Same as the profile described in Fig.~\ref{fig:J0823+6216} but for candidate ORCs.}
    \label{fig: cand orc profiles}
\end{figure*}

\end{appendix}

\end{document}